\documentclass[conference]{IEEEtran}
\IEEEoverridecommandlockouts
\usepackage{cite}
\usepackage[T1]{fontenc}
\usepackage{aecompl}
\usepackage[numbers,sort&compress]{natbib}
\usepackage{amsmath,amssymb,amsfonts}
\usepackage{float}
\usepackage{mathrsfs}
\usepackage{booktabs}
\makeatletter
\renewcommand{\maketag@@@}[1]{\hbox{\m@th\normalsize\normalfont#1}}%
\makeatother
\usepackage{algpseudocode}
\usepackage{graphicx}
\usepackage{caption}
\allowdisplaybreaks[4]
\usepackage{subfigure}
\usepackage{textcomp}
\usepackage{xcolor}
\usepackage[ruled,linesnumbered]{algorithm2e}
\makeatletter
\newcommand{\removelatexerror}{\let\@latex@error\@gobble}
\makeatother
\usepackage{
    caption}
\usepackage{float}
\usepackage{graphicx}
\usepackage{textcomp}
\usepackage{xcolor}
\usepackage{bm}
\usepackage{amsmath}
\usepackage{amsthm}
\usepackage{setspace}
\usepackage{geometry}
\usepackage{setspace}
\allowdisplaybreaks[4]
\newtheorem{theorem}{Theorem}
\newtheorem{corollary}{Corollary}
\newtheorem{lemma}{Lemma}

\def\BibTeX{{\rm B\kern-.05em{\sc i\kern-.025em b}\kern-.08em
    T\kern-.1667em\lower.7ex\hbox{E}\kern-.125emX}}
\def\d{\,\mathrm{d}}
\begin{document}
\title{Enhancing xURLLC with RSMA-Assisted Massive-MIMO Networks: Performance Analysis and Optimization
}
\author{Yuang~Chen,~\IEEEmembership{Student Member, IEEE}, Hancheng~Lu,~\IEEEmembership{Senior Member, IEEE}, Chenwu~Zhang, \\ Yansha~Deng,~\IEEEmembership{Senior Member, IEEE}, and Arumugam~Nallanathan,~\IEEEmembership{Fellow, IEEE}
\thanks{\setlength{\baselineskip}{1.2\baselineskip} Yuang Chen, Hancheng Lu, and Chengwu Zhang are with the CAS Key Laboratory of Wireless-Optical Communications, School of Information Science and Technology, University of Science and Technology of China, Hefei 230027, China (email: yuangchen21@mail.ustc.edu.cn; hclu@ustc.edu.cn; cwzhang@mail.ustc.edu.cn). Yansha~Deng is with King's College London (email: yansha.deng@kcl.ac.uk); Arumugam~Nallanathan is with Queen Mary University of London (email: a.nallanathan@qmul.ac.uk)).}}

\maketitle{}

\begin{abstract}
Massive interconnection has sparked people's envisioning for next-generation ultra-reliable and low-latency communications (xURLLC), prompting the design of customized next-generation advanced transceivers (NGAT). Rate-splitting multiple access (RSMA) has emerged as a pivotal technology for NGAT design, given its robustness to imperfect channel state information (CSI) and resilience to quality of service (QoS). Additionally, xURLLC urgently appeals to large-scale access techniques, thus massive multiple-input multiple-output (mMIMO) is anticipated to integrate with RSMA to enhance xURLLC. In this paper, we develop an innovative RSMA-assisted massive-MIMO xURLLC (RSMA-mMIMO-xURLLC) network architecture tailored to accommodate xURLLC's critical QoS constraints in finite blocklength (FBL) regimes. Leveraging uplink pilot training under imperfect CSI at the transmitter, we estimate channel gains and customize linear precoders for efficient downlink short-packet data transmission. Subsequently, we formulate a joint rate-splitting, beamforming, and transmit antenna selection optimization problem to maximize the total effective transmission rate (ETR). Addressing this multi-variable coupled non-convex problem, we decompose it into three corresponding subproblems and propose a low-complexity joint iterative algorithm for efficient optimization. Extensive simulations substantiate that compared with non-orthogonal multiple access (NOMA) and space division multiple access (SDMA), the developed architecture improves the total ETR by $15.3\%$ and $41.91\%$, respectively, as well as accommodates larger-scale access.
\end{abstract}

\begin{IEEEkeywords}
Next-generation ultra-reliable low-latency communications (xURLLC), rate-splitting multiple access (RSMA), massive multiple-input multiple-output (mMIMO), finite blocklength (FBL).
\end{IEEEkeywords}

\vspace{-1.2em}
\section{Introduction}
\vspace{-0.2em}
\subsection{Background}
\vspace{-0.3em}

\par The rapid evolution of the sixth-generation wireless networks (6G) has propelled the advent of the internet-of-everything (IoE) and indisputably catalyzed the advancement of ultra-reliable and low-latency communications (URLLC) \cite{park2022extreme, she2021tutorial, 10382447, 10355071, chen2023streaming}. URLLC endeavors to accomplish scalability and ultra-reliability with low latency \cite{park2022extreme, she2021tutorial}, prompting the continuous emergence of high-stack control and mission-critical applications across vertical sectors, such as extended reality, brain-machine interfaces, robotic control, and industrial automation \cite{park2022extreme,she2021tutorial,10382447,10355071}. These applications have provoked unprecedented demands for stringent quality-of-service (QoS), fueling the anticipation among stakeholders for the next-generation URLLC (xURLLC). To this end, the design of customized next-generation advanced transceivers (NGAT) is pivotal for implementing xURLLC, which necessitates exceptional scalability, robust multiple access, flexible interference management, elastic low latency, and impeccable reliability \cite{clerckx2023primer,mao2022rate,10382447,10355071}.

\par The predominant roadmap to fulfill low latency for xURLLC involves orchestrating extensive short-packet data transmissions within dynamic wireless networks \cite{khan2017wirelessly}. Finite blocklength (FBL) is essential to guarantee compliance with low-latency prerequisites \cite{she2021tutorial,10382447,10355071}, rendering Shannon's capacity theory inadequate due to non-negligible decoding error probability (DEP) in FBL regimes \cite{polyanskiy2010channel, 10382447,10355071}. To rectify this theoretical deficiency, finite blocklength coding theory has emerged, revealing the maximum achievable data rate for short-packet data transmissions while highlighting the intricate tradeoff between DEP and target latency. This tradeoff is not solely determined by blocklength but is also impacted by transmit power, transmission rate, channel state information (CSI), antenna amounts, and device scale \cite{she2021tutorial,10382447}. To meet xURLLC's QoS requirements, there arises the imminent necessity for innovative multi-access techniques to customize xURLLC's NGAT, which must deliver higher spectral efficiency (SE), greater energy efficiency (EE), more flexible interference management, wider coverage, and larger effective transmission rates (ETR) \cite{clerckx2023primer,mao2022rate,10382447,she2021tutorial,park2022extreme}.

\par Rate-splitting multiple access (RSMA) has emerged as a transformative multi-access technique, revolutionizing the architectural design and bolstering massive access, non-orthogonal transmission, and flexible interference management for 6G \cite{clerckx2023primer,mao2022rate,clerckx2024multiple}. It introduces a promising paradigm at the physical layer, expected to be exploited in the design of NGAT for xURLLC \cite{clerckx2023primer,mao2022rate,10382447,park2022extreme}. %It introduces a promising paradigm at the physical layer, poised to significantly enhance QoS for xURLLC \cite{clerckx2023primer,mao2022rate,10382447,park2022extreme}.
The fundamental principle of RSMA involves splitting the intended message for receivers into common and private parts, combining common parts into a super common message encoded using a shared codebook while encoding each receiver's private part independently using a distinct private codebook \cite{clerckx2023primer,mao2022rate}. These streams are transmitted leveraging linear precoding, enabling non-orthogonal transmission within the same time-frequency resources \cite{clerckx2023primer,mao2022rate,clerckx2024multiple}. By exploiting rate-splitting, RSMA can \emph{\textbf{partially decode the interference and partially treat the interference as noise}} \cite{clerckx2023primer,mao2022rate}, striking a balance between fully decoding interference and treating interference as noise \cite{clerckx2023primer,mao2022rate}. Therefore, RSMA stands as a pivotal technique for advancing xURLLC.

\vspace{-0.5em}

\subsection{Research Motivations and Challenges}

\vspace{-0.3em}

\par Although RSMA has demonstrated its potential to enhance QoS in wireless networks \cite{xu2022rate, xu2022max, pala2023spectral}, the research centered on the integration of RSMA into xURLLC remains in its nascent stages, accompanied by numerous unprecedented challenges. On the one hand, RSMA's capability to facilitate flexible interference management for receivers involves cumbersome message splitting and combining processes, which come at the expense of increasing framework complexity, posing formidable challenges for seamlessly integrating with other advanced technologies, like massive-MIMO \cite{10197580}, grant-free access \cite{10016266}, and FBL regime \cite{clerckx2023primer}. On the other hand, the optimal rate-splitting and beamforming schemes in complex wireless environments, especially under the FBL and massive-MIMO scenarios, are no simple feat \cite{clerckx2023primer,mao2022rate}. Additionally, for user terminals to successfully decode and eliminate common and private messages at receivers leveraging successive interference cancellation (SIC) techniques, user terminals must possess knowledge of RSMA splitting and combining short-packet messages at the transmitter. However, since the extremely complex integration architecture of xURLLC and RSMA, this remains an unresolved challenge.

\vspace{-0.1em}

\par The predominant claim of xURLLC center on supporting the rapidly proliferating device connectivity while meeting stringent QoS demands. Despite RSMA's potential to enhance xURLLC \cite{pala2023spectral, 10355071, 10382447}, addressing massive access remains a imminent challenge \cite{clerckx2023primer,mao2022rate}. Massive MIMO (mMIMO) emerges as a promising solution, boasting advantages like enhanced spectral efficiency (SE) and energy efficiency (EE) by over tenfold and a hundredfold, respectively \cite{marzetta2010noncooperative,6798744}. However, integrating RSMA with mMIMO is confronted with several unexploited challenges. Firstly, RSMA-assisted mMIMO (RSMA-mMIMO) systems rely heavily on spatial multiplexing, demanding accurate channel knowledge for the transceiver's reliable uplink and downlink transmissions. Secondly, effective beamforming and interference management of RSMA-mMIMO systems hinges on precise CSIT, which is typically unavailable in practical scenarios, leading to potential multi-user interference and compromised latency and reliability. Finally, transmit antenna selection of RSMA-mMIMO systems demands judicious deliberation to avoid escalating hardware costs, power consumption \cite{rusek2012scaling}, signal processing complexity \cite{bjornson2014massive}, compromised SE \cite{marzetta2010noncooperative}, and interference and DEP sensitivity \cite{larsson2014massive}. Therefore, tailoring an innovative RSMA-mMIMO system for xURLLC necessitates meticulous study to imperfect CSIT in channel estimation, interference management, transmit antenna selection, and resource allocation schemes.

\vspace{-0.5em}

\subsection{Related Work}

\vspace{-0.5em}
\par Sate-of-the-art research endeavors have extensively focused on optimizing short-packet transmission for xURLLC \cite{10382447,10355071,lin2022joint,ke2023next}. Our prior work has developed an xURLLC-enabled multi-user MIMO network to minimize the upper-bounded statistical delay violation probability by jointly optimizing the pilot length and precoding matrix \cite{10382447}. FBL-based short packets with sporadic arrival traffic necessitate tail distribution exploration. For this reason, Y. Chen \emph{et al.} in \cite{10355071} have proposed a non-orthogonal multiple access (NOMA)-assisted uplink xURLLC network that incorporates a theoretical framework for tail analysis. Joint design algorithms for channel training and data transmission have presented formidable challenges in rigorous reliability and FBL regimes, leading to proposing a low-complexity joint design framework for MIMO-enabled URLLC networks \cite{lin2022joint}. Moreover, a unified semi-blind detection framework has been proposed to facilitate massive access for xURLLC, enhancing scalability, reliability, and latency tradeoffs \cite{ke2023next}.

\par RSMA has emerged as a pioneering multi-access technology to enhance xURLLC's QoS provisioning. Research has thoroughly explored the intricate relationship between the sum rate and blocklength of RSMA-assisted URLLC, addressing various network loads and user deployment scenarios \cite{xu2022rate}. Notably, RSMA endows with the capability to guarantee user fairness in both underloaded and overloaded deployment paradigms \cite{zhang2019cooperative}. Authors in \cite{xu2022max} have proposed a novel non-cooperative/cooperative multi-group multicast deployment model to maximize the minimum group rate for URLLC networks. To fulfill xURLLC's stringent QoS requirements, authors in \cite{pala2023spectral} have investigated a multi-reconfigurable intelligent surface (RIS)-assisted RSMA architecture, aiming to prompt enhanced SE and maximize the sum throughput. Additionally, an innovative cooperative RSMA framework in a two-layer heterogeneous network has been proposed, which leverages users with favorable channel quality as full-duplex relays to assist users with weaker signals \cite{10233866}.

\vspace{-0.1em}

\par Extensive studies have explored performance enhancements in RSMA-mMIMO networks \cite{10197580, dizdar2021rate,jiang2023rate,zhang2019cooperative}. A novel RSMA-based massive-MIMO system has been proposed to analyze the implications of channel aging, spatial correlation, and Rician factor \cite{10197580}. The delay between CSI acquisition and data transmission stands as the primary factor contributing to imperfect CSIT. Authors in \cite{dizdar2021rate} have derived the lower bound of the ergodic sum rate and investigated the optimal power allocation for common and private streams. To address concerns about long-term electromagnetic (EM) radiation exposure, RSMA has been applied to enhance SE and EE of uplink massive MIMO while rigorously constraining EM exposure \cite{jiang2023rate}. Leveraging the polarization domain for additional degrees of freedom (DoF), authors in \cite{zhang2019cooperative} have improved the successive interference cancellation (SIC) performance and developed three dual-polarized downlink transmission strategies for RSMA-mMIMO systems.

\vspace{-0.5em}

\subsection{Main Contributions}

\vspace{-0.3em}

\par To effectively address the aforementioned challenges, we propose an innovative RSMA-assisted massive-MIMO xURLLC (RSMA-mMIMO-xURLLC) network, which has been demonstrated unequivocally furnish exceptional QoS provisioning for xURLLC. We adopt the ETR as a performance indicator to evaluate the relationship between the achievable data rate and xURLLC's QoS requirements. Subsequently, we formulate a joint rate-splitting, beamforming, and transmit antenna selection optimization problem to maximize the total ETR. To efficiently solve this problem, we propose a novel joint optimization mechanism to decompose the original problem into three corresponding subproblems. Furthermore, a low-complexity joint iterative optimization algorithm is designed to efficiently alternate the optimization of power allocation, rate-splitting, and transmit antenna selection, respectively. Exhaustive experiments substantiate the extraordinary performance of the developed RSMA-mMIMO-xURlLC network compared to several state-of-the-art multi-access frameworks. The main contributions of this paper are summarized as follows:

\vspace{-0.1em}

\begin{itemize}
  \item We develop an innovative RSMA-mMIMO-xURLLC network architecture, where the transceiver is equipped with multiple antennas to fully exploit the potential of RSMA-mMIMO systems. The imperfect CSIT and the FBL regime have been accounted for meticulously accommodating xURLLC's critical QoS requirements. Moreover, uplink pilot training has been undertaken to estimate channel gains, and the linear precoder has been tailored for efficient downlink short-packet data transmission.

  \item We derive closed-form expressions for the signal-to-interference-plus-noise ratios (SINRs) of common and private streams, as well as the corresponding channel capacity. Subsequently, we formulate a joint rate-splitting, beamforming, and transmit antenna selection optimization problem to maximize the total ETR. To effectively overcome this problem, we reformulate it by profoundly exploring the relationships between SINRs of common and private streams with antenna amounts, large-scale fading, transmit power, blocklength, and channel estimation quality (CEQ).

  \item We decompose the reformulated problem into three corresponding subproblems. In particular, the successive convex approximation (SCA)-based algorithm is proposed to solve the power allocation subproblem, while the rate-splitting subproblem is effectively addressed by interior-point methods. Furthermore, the transmit antenna selection subproblem is equivalently transformed into a one-dimensional integer search problem. Ultimately, a low-complexity joint iterative optimization algorithm is proposed to efficiently alternate the optimization of these three subproblems.

  \item Extensive simulations substantiate the optimality and convergence performance of the proposed joint iterative optimization algorithm. Additionally, compared with the state-of-the-art multi-access schemes, including NOMA, and space division multiple access (SDMA), the extraordinary performance of our developed RSMA-mMIMO-xURLLC network architecture can be demonstrated in terms of ETR, reliability, latency, and robustness for xURLLC.
\end{itemize}

\vspace{-0.2em}

\par The remainder of this paper is organized as follows. In Sec. II, the RSMA-mMIMO-xURLLC network architecture is investigated. Sec. III encompasses the problem formulation and solution. Section IV gives the performance evaluation and analysis, followed by the conclusion and future outlook.

\section{RSMA-mMIMO-xURLLC Network Architecture}
\vspace{-0.4em}
\par
\begin{figure}[htbp]
\vspace{-0.7em}
\centering
\includegraphics[scale=0.55]{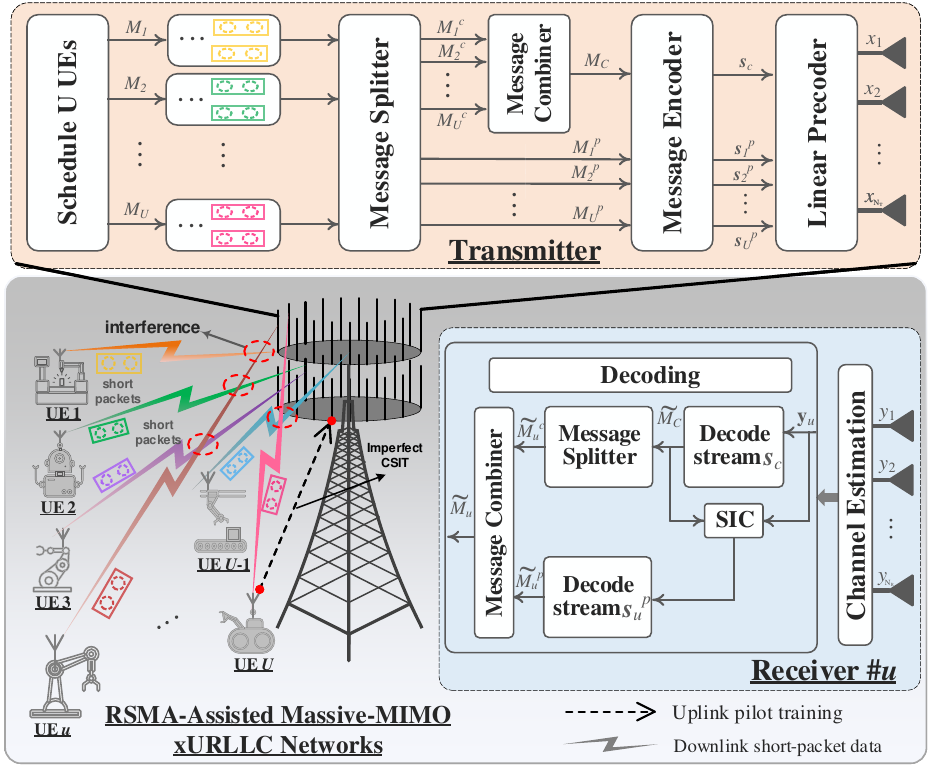}
\vspace{-0.2em}
\caption{The RSMA-assisted massive-MIMO xURLLC network architecture.}
\label{fig1}
\vspace{-1.2em}
\end{figure}

 As illustrated in Fig. \ref{fig1}, we examine a downlink RSMA-mMIMO-xURLLC network architecture, where a base station (BS) equipped with $N_{T}$ antennas concurrently serves $U$ user equipments (UEs) equipped with $N_{R}$ antennas ($N_{T} \gg N_{R}$). Each UE $u$ requests message $M_{u}$ from BS, where $u \in \mathcal{U}\triangleq\left\{1,\cdots,U\right\}$. Following RSMA principles \cite{clerckx2023primer,mao2022rate}, $M_{u}$ is split into a common message $M_{u}^{c}$ and a private message $M_{u}^{p}$. All common messages $\left\{M_{1}^{c},\cdots,M_{U}^{c}\right\}$ are combined into a super-common message $M_{c}$, encoded into a common stream $s_{c}$. The codebook of $s_{c}$ is shared with all receivers to ensure successful decoding of $s_{c}$. Conversely, each $M_{u}^{p}$ is independently encoded into a private stream $s_{u}^{p}$, with its codebook shared solely with its designated receiver $u \in \mathcal{U}$. Subsequently, streams $\boldsymbol{\mathrm{s}}\triangleq \left[s_{1}^{p},\cdots,s_{U}^{p},s_{c}\right]^{T} \in \mathbb{C}^{(U+1)\times 1}$, $\mathbb{E}\left[\boldsymbol{\mathrm{s}}\boldsymbol{\mathrm{s}}^{\mathrm{H}}\right] = \boldsymbol{\mathrm{I}}$ are precoded into transmitted signals $\boldsymbol{\mathrm{x}} \triangleq \left[x_{1},\cdots,x_{N_{T}}\right]^{T}$ through linear precoding, where $x_{i}$ denotes the signal corresponding to the $i$-th transmit antenna ($i = 1,\cdots,N_{T}$).

\vspace{-0.7em}

\subsection{Critical QoS Provisioning Requirements for xURLLC}

\vspace{-0.3em}

\par To fulfill the extremely low latency for xURLLC, short-packet data transmission under FBL regimes is mandated to guarantee compliance with low-latency prerequisites \cite{yang2014quasi,polyanskiy2010channel,10382447}. In this context, the maximum achievable rate for multi-antenna receivers can be given by

\vspace{-1.2em}

\begin{small}
\begin{equation}\label{e1}
   \!\!\!\!\! R\!\left(N_{d},\boldsymbol{\mathrm{\Gamma}},\varepsilon\right) \! \approx \! \sum\limits_{j=1}^{N_{R}} \!\!\bigg(\!\! \log_{2}\!\left(\! 1 \!+\! \Gamma^{(j)}\!\right) \!-\! \sqrt{\frac{\mathcal{V}\left(\Gamma^{(j)}\right)}{N_{d}}}\frac{Q^{-1}\left(\varepsilon\right)}{\ln 2}\!\bigg),
   \vspace{-0.4em}
\end{equation}
\end{small}
where $\Gamma^{(j)}$ represents the SINR received by the $j$-th antenna of the receiver, $N_{d}$ indicates the allocated channel uses (CUs) for xURLLC's short-packet data transmission, $\varepsilon$ represents DEP, $Q^{-1}(x)$ symbolizes the inverse of the Q-function, and $\mathcal{V}\left(\Gamma^{(j)}\right) = 1 - \left(1 + \Gamma^{(j)}\right)^{-2}$ characterizes the channel dispersion. Then, the expression of DEP can be expressed as
\vspace{-0.3em}
\begin{equation}\label{e2}
   \varepsilon\left(N_{d},\boldsymbol{\mathrm{\Gamma}},R\right) = Q\big(g\left(N_{d},\boldsymbol{\mathrm{\Gamma}},R\right)\big),
   \vspace{-0.4em}
\end{equation}
where
\vspace{-0.4em}
\begin{small}
\begin{equation*}
   g\!\left(N_{d},\boldsymbol{\mathrm{\Gamma}},R\right) \!=\! \ln 2 \bigg(\!\sum\limits_{j = 1}^{N_{R}}\!\log_{2}\!\!\big(1 + \Gamma^{(j)}\big)-R\!\bigg)\!\bigg(\!\sum\limits_{j = 1}^{N_{R}}\!\sqrt{\frac{\mathcal{V}\left(\Gamma^{(j)}\right)}{N_{d}}}\!\bigg)^{\!-1}\!\!\!.
\end{equation*}
\end{small}

\vspace{-0.7em}

\par Let $\!D_{th}\!$ and $\!\varepsilon_{th}\!$ characterize xURLLC's QoS requirements for latency and reliability, respectively \cite{10382447,10355071,she2021tutorial}. Given the system bandwidth $B_{tot}$, the symbol duration and the total CUs can be represented by $T_{f} \!=\! 1/B_{tot}$ and $N_{tot} \!=\! D_{th}/T_{f}$, respectively. Under FBL regimes, the short-packet data communications must be accomplished within $N_{d}$ CUs. As a result, the latency constraint of xURLLC is formulated as  $N_{d} \!\leq\! N_{tot} \!-\! N_{p}$, where $\!N_{p}\!$ is the CUs allocated for pilot training, and the reliability constraint can be formulated as $\varepsilon\left(N_{d},\boldsymbol{\mathrm{\Gamma}},R\right) \leq \varepsilon_{th}$.

\vspace{-0.6em}

\subsection{Uplink Pilot Training for Channel Estimations}

\vspace{-0.4em}

\par Let $\boldsymbol{\varphi} = \left[\varphi_{1},\cdots,\varphi_{N_{p}}\right] \in \mathbb{C}^{1\times N_{p}}$ denote the orthogonal pilot sequence satisfying $\|\boldsymbol{\varphi}\|^{2} = 1$, where operator $\|\cdot\|$ indicates the Euclidean norm ($N_{p} > N_{R}$). The uplink pilot signal from UE $u$ to the BS can be represented as $\boldsymbol{\mathrm{x}}_{u}^{(\mathrm{p})} = \sqrt{N_{p}}\boldsymbol{\varphi}$ \cite{marzetta2016fundamentals}. Let $\rho_{p}$ represent the transmit power for uplink pilot training, then the received pilot signal of the $i$-th receive antenna is given by

\vspace{-1em}

\begin{equation}\label{e3}
   \boldsymbol{\mathrm{Y}}_{u,i}^{(\mathrm{p})} = \sqrt{N_{p}}\boldsymbol{\mathrm{g}}_{u,i}\boldsymbol{\mathrm{x}}_{u}^{(\mathrm{p})} + \boldsymbol{\mathrm{N}}^{(\mathrm{p})} = \sqrt{N_{p}\rho_{p}} \boldsymbol{\mathrm{g}}_{u,i}\boldsymbol{\varphi} + \boldsymbol{\mathrm{N}}^{(\mathrm{p})},
\end{equation}
where $\boldsymbol{\mathrm{N}}^{(\mathrm{p})} \in \mathbb{C}^{N_{R} \times N_{p}}$ indicates the additive white Gaussian noise (AWGN) matrix, $\boldsymbol{\mathrm{g}}_{u,i} = \sqrt{\kappa_{u}} \boldsymbol{\mathrm{h}}_{u,i}$ denotes the channel gain between the UE $u$ and the $i$-th transmit antenna over the transmitter, where $\kappa_{u} = [\lambda/(4\pi d_{u})]^{2}$ and $\lambda$ are the large-scale fading coefficient and the wavelength, respectively. While $d_{u}$ and $\boldsymbol{\mathrm{h}}_{u,i} \in \mathbb{C}^{N_{R} \times 1}$ represents the distance and the small-scale fading coefficient between the UE $u$ and the $i$-th transmit antenna, respectively. The BS undertakes the de-spreading scheme to the received uplink pilot signal \cite{marzetta2016fundamentals}, which can be given by (\ref{e4}) after the de-spreading operation:
\vspace{-0.2em}
\begin{equation}\label{e4}
   \boldsymbol{\overline{\mathrm{y}}}_{u,i}^{(\mathrm{p})} = \boldsymbol{\mathrm{Y}}_{u,i}^{(\mathrm{p})}\boldsymbol{\varphi}^{\mathrm{H}} = \sqrt{N_{p}\rho_{p}} \boldsymbol{\mathrm{g}}_{u,i} + \boldsymbol{\overline{\mathrm{n}}}^{(\mathrm{p})}
\end{equation}
where each element of $\boldsymbol{\overline{\mathrm{n}}}^{(\mathrm{p})} \in \mathbb{R}^{N_{R}\times 1}$ follows the Gaussian distribution $\mathcal{N}(0,1)$. Let $\boldsymbol{\widehat{G}}_{u} = \left[\boldsymbol{\widehat{\mathrm{g}}}_{u,1},\cdots,\boldsymbol{\widehat{\mathrm{g}}}_{u,N_{T}}\right] \in \mathbb{C}^{N_{R}\times N_{T}}$ indicate the estimation of the channel gain matrix $\boldsymbol{G}_{u} = \left[\boldsymbol{\mathrm{g}}_{u,1},\cdots,\boldsymbol{\mathrm{g}}_{u,N_{T}}\right] \in \mathbb{C}^{N_{R}\times N_{T}}$. By exploiting the minimum mean-square error (MMSE) estimation, we have \cite{marzetta2016fundamentals}

\vspace{-0.6em}

\begin{equation}\label{e5}
   \boldsymbol{\widehat{\mathrm{g}}}_{u,i} = \mathbb{E}\left[\boldsymbol{\mathrm{g}}_{u,i}|\boldsymbol{\overline{\mathrm{y}}}_{u,i}^{(\mathrm{p})}\right] = \frac{\sqrt{N_{p}\rho_{p}}\kappa_{u}}{1 + N_{p}\rho_{p} \kappa_{u}} \boldsymbol{\overline{\mathrm{y}}}_{u,i}^{(\mathrm{p})}.
\end{equation}

\vspace{-0.5em}

\par By substituting (\ref{e4}) into (\ref{e5}), the estimated channel gain $\boldsymbol{\widehat{\mathrm{g}}}_{u,i}$ can be given as follows:

\vspace{-0.7em}

\begin{small}
\begin{equation}\label{e6}
   \boldsymbol{\widehat{\mathrm{g}}}_{u,i} = \frac{N_{p}\rho_{p}\kappa_{u}}{1 + N_{p}\rho_{p} \kappa_{u}}\boldsymbol{\mathrm{g}}_{u,i} + \frac{\sqrt{N_{p}\rho_{p}}\kappa_{u}}{1 + N_{p}\rho_{p} \kappa_{u}} \boldsymbol{\overline{\mathrm{n}}}^{(\mathrm{p})}
\end{equation}
\end{small}

\vspace{-1.0em}

\subsection{Precoder Design for Beamforming}

\vspace{-0.4em}
\par By leveraging RSMA technique \cite{clerckx2023primer,mao2022rate}, the transmitted signal $\boldsymbol{\mathrm{x}} \in \mathbb{C}^{N_{T}\times 1}$ can be given by

\vspace{-0.6em}

\begin{equation}\label{e7}
   \boldsymbol{\mathrm{x}} = \sqrt{\rho_{c}} \boldsymbol{\mathrm{w}}_{c} \boldsymbol{\mathrm{s}}_{c} + \theta \sum\limits_{u = 1}^{U} \sqrt{\rho_{p,u}} \boldsymbol{\mathrm{w}}_{p,u} \boldsymbol{\mathrm{s}}_{p,u},
   \vspace{-0.3em}
\end{equation}
where $\theta$ denotes the parameter that normalizes the total transmit power to $1$, $\rho_{c}$ and $\rho_{p,u}$ are the transmit power allocated for $\boldsymbol{\mathrm{s}}_{c}$ and $\boldsymbol{\mathrm{s}}_{p,u}$, respectively. While $\boldsymbol{\mathrm{w}}_{c} \in \mathbb{C}^{N_{T}\times 1}$ and $\boldsymbol{\mathrm{W}}_{p} = \left[\boldsymbol{\mathrm{w}}_{p,1},\cdots,\boldsymbol{\mathrm{w}}_{p,U}\right]\in\mathbb{C}^{N_{T}\times U}$ indicate the precoders designed for $\boldsymbol{\mathrm{s}}_{c}$ and $\left[\boldsymbol{\mathrm{s}}_{p,1},\cdots,\boldsymbol{\mathrm{s}}_{p,U}\right]$, respectively. By exploiting the zero-forcing (ZF) precoding scheme, $\boldsymbol{\mathrm{W}}_{p}$ can be denoted as
\vspace{-0.2em}
\begin{equation}\label{e8}
   \boldsymbol{\mathrm{W}}_{p} = \sqrt{N_{T} - N_{R}} \boldsymbol{\mathrm{Z}}^{\ast}\left(\boldsymbol{\mathrm{Z}}^{\mathrm{T}}\boldsymbol{\mathrm{Z}}^{\ast}\right)^{-1},
   \vspace{-0.2em}
\end{equation}
where $\boldsymbol{\mathrm{Z}} \in \mathbb{C}^{N_{T} \times U}$, and the elements of $\boldsymbol{\mathrm{Z}}$ are i.i.d. and follows $\mathcal{CN}(0,1)$. From $(\ref{e8})$, the normalized parameter $\theta$ can be obtained as follows \cite{raeesi2018performance}:
\vspace{-0.5em}
\begin{small}
\begin{equation}\label{e9}
   \theta = \left(\mathbb{E}\left[\mathrm{Tr}\left((\boldsymbol{\mathrm{W}}_{p})^{\mathrm{H}} \boldsymbol{\mathrm{W}}_{p} \right)\right]\right)^{-\frac{1}{2}} = \sqrt{\frac{\left(N_{T} - N_{R}\right) N_{p}\rho_{p}}{N_{R}\left(N_{p}\rho_{p}+1\right)}}.
\end{equation}
\end{small}
\vspace{-0.4em}

\par According to \cite{dai2016rate, clerckx2023primer,mao2022rate}, the precoder $\boldsymbol{\mathrm{w}}_{c}$ for the common stream $\boldsymbol{\mathrm{s}}_{c}$ can be given as follow:

\vspace{-0.6em}

\begin{equation}\label{e10}
   \boldsymbol{\mathrm{w}}_{c} = \sum\limits_{u = 1}^{U} \xi_{u} \left(\boldsymbol{\widehat{\mathrm{g}}}_{u}\right)^{\mathrm{H}},
   \vspace{-0.5em}
\end{equation}
where $\boldsymbol{\widehat{\mathrm{g}}}_{u}$ denotes the row vector of $\boldsymbol{\widehat{G}}_{u}$. Remarkably, since the distances of all antennas on each UE $u$ are very close, we can reasonably have $\boldsymbol{\mathrm{g}}_{u}^{(j)} \approx \boldsymbol{\mathrm{g}}_{u}^{(j^{\prime})} = \boldsymbol{\mathrm{g}}_{u}$ and $\boldsymbol{\widehat{\mathrm{g}}}_{u}^{(j)} \approx \boldsymbol{\widehat{\mathrm{g}}}_{u}^{(j^{\prime})} = \boldsymbol{\widehat{\mathrm{g}}}_{u}$ ($\forall j,j^{\prime} \in \left\{1,\cdots,N_{R}\right\}, j \neq j^{\prime}$), where $\boldsymbol{\mathrm{g}}_{u}^{(j)}$ and $\boldsymbol{\widehat{\mathrm{g}}}_{u}^{(j)}$ denote the $j$-th row vector of $\boldsymbol{G}_{u}$ and $\boldsymbol{\widehat{G}}_{u}$, respectively. And the expression of $\xi_{u}$ can be given by

\vspace{-0.5em}
\begin{small}
\begin{equation}\label{e11}
  \xi_{u} = 1\bigg/\sqrt{N_{T}\sum\limits_{k = 1}^{U}\frac{\pi_{u}(1 - \varpi_{u}^{2})}{\pi_{k}(1 - \varpi_{k}^{2})}},
\end{equation}
\end{small}
\vspace{-0.7em}
where
\vspace{-0.3em}
\begin{small}
\begin{equation}\label{e12}
  \varpi_{u} = \frac{\sqrt{N_{p}\rho_{p}}\kappa_{u}}{1 + N_{p}\rho_{p}\kappa_{u}},\ \pi_{u} = \rho_{c}\left(\sum\limits_{u=1}^{U}\rho_{p,u}|\boldsymbol{\mathrm{g}}_{u}\boldsymbol{\mathrm{w}}_{p,u}|^{2} + 1\right)^{-1},
\end{equation}
\end{small}
in which $\varpi_{u}$ indicates the channel estimation quality of receiver $u \in \mathcal{U}$. When $\varpi_{u} \rightarrow 0$, it indicates the perfect CSI estimation; while $\varpi_{u} \rightarrow 1$ denotes that the estimated channel is uncorrelated with the actual channel.

\vspace{-0.7em}

\subsection{Downlink Short-Packet Data Transmission}

\vspace{-0.2em}

\par The received signals corresponding to the $j$-th antenna over the receiver $u \in \mathcal{U}$ can be given by
\vspace{-0.2em}
\begin{small}
\begin{equation}\label{e13}
   \begin{aligned}
       & y_{u}^{(j)} = \boldsymbol{\mathrm{g}}_{u}^{(j)} \boldsymbol{\mathrm{x}} + w_{u}^{(j)} = \underbrace{\sqrt{\rho_{c}} \boldsymbol{\mathrm{g}}_{u}^{(j)} \boldsymbol{\mathrm{w}}_{c} \boldsymbol{\mathrm{s}}_{c}}_{\text{common message}} + w_{u}^{(j)} \\
       & + \theta \bigg(\underbrace{\sqrt{\rho_{p,u}} \boldsymbol{\mathrm{g}}_{u}^{(j)} \boldsymbol{\mathrm{w}}_{p,u} \boldsymbol{\mathrm{s}}_{p,u}}_{\text{private message for UE $u$}} + \underbrace{\sum\limits_{k = 1, k \neq u}^{U}\sqrt{\rho_{p,k}} \boldsymbol{\mathrm{g}}_{u}^{(j)} \boldsymbol{\mathrm{w}}_{p,k} \boldsymbol{\mathrm{s}}_{p,k}}_{\text{interference for UE $u$}}\bigg),
   \end{aligned}
\end{equation}
\end{small}
where $w_{u}^{(j)}$ denotes the AWNG noise received by the $j$-th antenna of UE $u \in \mathcal{U}$. Following RSMA principles \cite{dai2016rate, clerckx2023primer,mao2022rate}, we first decode the most important common message, which is interfered with by all UEs' private messages. Therefore, the SINR of the common message received by the $j$-th antenna on UE $u$ is given by

\vspace{-0.5em}

\begin{small}
\begin{equation}\label{e14}
   \Gamma_{c,u}^{(j)} = \frac{\mathrm{Var}\left[\sqrt{\rho_{c}} \boldsymbol{\mathrm{g}}_{u}^{(j)} \boldsymbol{\mathrm{w}}_{c} \boldsymbol{\mathrm{s}}_{c}\right]}{\sum_{k = 1}^{U}\mathrm{Var}\left[\sqrt{\rho_{p,k}} \boldsymbol{\mathrm{g}}_{u}^{(j)} \boldsymbol{\mathrm{w}}_{p,k} \boldsymbol{\mathrm{s}}_{p,k}\right] + 1}.
\end{equation}
\end{small}

\vspace{-0.2em}

\par Subsequently, we decode the private message $\boldsymbol{\mathrm{s}}_{p,u}$. If $\boldsymbol{\mathrm{s}}_{c}$ can be decoded successfully, then $\boldsymbol{\mathrm{s}}_{p,u}$ is interfered with by private messages from other UEs, and the corresponding SINR can be given by

\vspace{-1.2em}

\begin{small}
\begin{equation}\label{e15}
  \Gamma_{p,u}^{(j)} = \frac{\mathrm{Var}\left[\sqrt{\rho_{p,u}} \boldsymbol{\mathrm{g}}_{u}^{(j)} \boldsymbol{\mathrm{w}}_{p,u} \boldsymbol{\mathrm{s}}_{p,u}\right]}{\sum_{k = 1, k \neq u}^{U}\mathrm{Var}\left[\sqrt{\rho_{p,k}} \boldsymbol{\mathrm{g}}_{u}^{(j)} \boldsymbol{\mathrm{w}}_{p,k} \boldsymbol{\mathrm{s}}_{p,k}\right] + 1}.
\end{equation}
\end{small}

\vspace{-0.2em}

\par If receiver $u$ fails to decode $\boldsymbol{\mathrm{s}}_{c}$, the decoding of $\boldsymbol{\mathrm{s}}_{p,u}$ is interfered with by both $\boldsymbol{\mathrm{s}}_{c}$ and the private messages of other UEs. In this case, the corresponding SINR is given by

\vspace{-0.9em}

\begin{small}
\begin{equation}\label{e16}
  \widehat{\Gamma}_{p,u}^{(j)} = \frac{\mathrm{Var}\!\left[\sqrt{\rho_{p,u}} \boldsymbol{\mathrm{g}}_{u}^{(j)} \boldsymbol{\mathrm{w}}_{p,u} \boldsymbol{\mathrm{s}}_{p,u}\right]}{\mathrm{Var}\!\left[\sqrt{\rho_{c}} \boldsymbol{\mathrm{g}}_{u}^{(j)} \boldsymbol{\mathrm{w}}_{c} \boldsymbol{\mathrm{s}}_{c}\right] + \!\!\!\!\! \sum\limits_{k = 1, k \neq u}^{U}\!\!\mathrm{Var}\!\left[\sqrt{\rho_{p,k}} \boldsymbol{\mathrm{g}}_{u}^{(j)} \boldsymbol{\mathrm{w}}_{p,k} \boldsymbol{\mathrm{s}}_{p,k}\right] \!+\! 1}.
\end{equation}
\end{small}

\vspace{-0.8em}

\section{Problem Formulation And Solution}

\vspace{-0.2em}

\subsection{Problem Formulation}

\vspace{-0.2em}

\par Under FBL regimes \cite{yang2014quasi,polyanskiy2010channel,10382447}, the imperfect SIC caused by non-vanishing DEP need to be incorporated. The DEP of receiver $u$ decoding common stream $\boldsymbol{\mathrm{s}}_{c}$ can be expressed as
\vspace{-0.3em}
\begin{equation}\label{e17}
   \varepsilon_{c,u}\!\left(N_{d},\boldsymbol{\mathrm{\Gamma}}_{c,u},\mathcal{R}_{c}\right) = Q\big(g\left(N_{d},\boldsymbol{\mathrm{\Gamma}}_{c,u},\mathcal{R}_{c}\right)\big),
   \vspace{-0.5em}
\end{equation}
where $\mathcal{R}_{c}$ indicates the transmission rate of $\boldsymbol{\mathrm{s}}_{c}$, and $\boldsymbol{\mathrm{\Gamma}}_{c,u} = \left[\Gamma_{c,u}^{(1)},\cdots,\Gamma_{c,u}^{(N_{R})}\right]^{T}$. To guarantee all receivers can successfully decode $\boldsymbol{\mathrm{s}}_{c}$, $\mathcal{R}_{c}$ must fulfill the constraints as follows:

\vspace{-1.5em}

\begin{equation}\label{e18}
   \!\!\!\!\!\! \mathcal{R}_{c} \!\leq\! \min_{u \in \mathcal{U}}\big\{\! C_{c,u}\!\left(N_{d},\boldsymbol{\mathrm{\Gamma}}_{\!c,u},\varepsilon\right)\!\big\} \! \leq \!\min_{u \in \mathcal{U}}\big\{\! C_{c,u}\!\left(N_{d},\boldsymbol{\mathrm{\Gamma}}_{\!c,u},\varepsilon_{th}\!\right)\!\big\},\!\!
   \vspace{-0.5em}
\end{equation}
where $C_{c,u}\!\left(N_{d},\boldsymbol{\mathrm{\Gamma}}_{c,u},\varepsilon\right)$ denotes the maximum achievable rate when decoding $\boldsymbol{\mathrm{s}}_{c}$, and it can be given by

\vspace{-1.4em}

\begin{small}
\begin{equation}\label{e19}
     \!\!\! C_{c,u}\!\big(N_{d},\boldsymbol{\mathrm{\Gamma}}_{c,u},\varepsilon\big) \!=\! \sum\limits_{j = 1}^{N_{R}} \! \bigg(\!\!\log_{2}\!\left(\! 1 \!+ \Gamma_{c,u}^{(j)}\right) \!-\! \sqrt{\!\!\frac{\mathcal{V}\left(\Gamma_{c,u}^{(j)}\right)}{N_{d}}}\frac{Q^{-1}\left(\varepsilon\right)}{\ln 2}\!\bigg).
\end{equation}
\end{small}

\par Let $\mathcal{R}_{c,u}$ denote the common rate allocated to receiver $u \in \mathcal{U}$, it satisfies that

\vspace{-1em}
\begin{small}
\begin{equation}\label{e20}
   \sum\limits_{u = 1}^{U} \mathcal{R}_{c,u} \leq \mathcal{R}_{c}.
\end{equation}
\end{small}

\vspace{-0.5em}

\par Combining (\ref{e17})-(\ref{e20}), the ETR of the common stream $\boldsymbol{\mathrm{s}}_{c}$ can be given by

\vspace{-0.5em}

\begin{equation}\label{e21}
   \widetilde{\mathcal{R}}_{c,u} = \mathcal{R}_{c,u}\cdot \left(1 - \varepsilon_{c,u}\big(N_{d},\boldsymbol{\mathrm{\Gamma}}_{c,u},\mathcal{R}_{c}\big)\right).
\end{equation}

\par If receiver $u\in\mathcal{U}$ successfully decodes $\boldsymbol{\mathrm{s}}_{c}$, the DEP of $\boldsymbol{\mathrm{s}}_{p,u}$ can be represented by $\varepsilon_{p,u}\!\left(N_{d},\boldsymbol{\mathrm{\Gamma}}_{p,u},\mathcal{R}_{p,u}\right)$. If receiver $u\in\mathcal{U}$ fails to decode $\boldsymbol{\mathrm{s}}_{c}$, the DEP of $\boldsymbol{\mathrm{s}}_{p,u}$ can be denoted as $\widehat{\varepsilon}_{p,u}\!\left(N_{d},\boldsymbol{\mathrm{\widehat{\Gamma}}}_{p,u},\mathcal{R}_{p,u}\right)$, where $\boldsymbol{\mathrm{\Gamma}}_{p,u} = \left[\Gamma_{p,u}^{(1)},\cdots,\Gamma_{p,u}^{(N_{R})}\right]^{T}$ and $\boldsymbol{\mathrm{\widehat{\Gamma}}}_{p,u} = \left[\widehat{\Gamma}_{p,u}^{(1)},\cdots,\widehat{\Gamma}_{p,u}^{(N_{R})}\right]^{T}$. $\mathcal{R}_{p,u}$ is the transmission rate allocated to $\boldsymbol{\mathrm{s}}_{p,k}$. Therefore, the ETR of $\boldsymbol{\mathrm{s}}_{p,u}$ is given by

\vspace{-1.2em}

\begin{equation}\label{e22}
   \begin{aligned}
     \widetilde{\mathcal{R}}_{p,u} & = \mathcal{R}_{p,u}\cdot \big(1 - \varepsilon_{c,u}\widehat{\varepsilon}_{p,u} - \left(1 - \varepsilon_{c,u}\right)\varepsilon_{p,u}\big)\\
      & \approx \mathcal{R}_{p,u}\cdot\big(1 - \varepsilon_{p,u}\!\left(N_{d},\boldsymbol{\mathrm{\Gamma}}_{p,u},\mathcal{R}_{p,u}\right)\big).
   \end{aligned}
\end{equation}

\vspace{-0.5em}

\par Subsequently, we formulate a joint rate splitting, power allocation, transmit antenna selection, pilot length, and blocklength optimization problem to maximize the total ETR while guaranteeing QoS requirements for the developed RSMA-mMIMO-xURLLC network architecture, as follows:

\vspace{-1.5em}

\begin{subequations}
   \begin{align}
      \mathcal{P}1:&\ \max_{\{N_{p},N_{d},N_{T},\boldsymbol{\mathcal{R}},\boldsymbol{\rho}\}} \sum\limits_{u \in \mathcal{U}} \bigg(\mathcal{R}_{c,u}\cdot \left(1 - \varepsilon_{c,u}\big(N_{d},\boldsymbol{\mathrm{\Gamma}}_{c,u},\mathcal{R}_{c}\big)\right) \nonumber \\
      & \quad \quad \quad + \mathcal{R}_{p,u}\cdot\big(1 - \varepsilon_{p,u}\!\left(N_{d},\boldsymbol{\mathrm{\Gamma}}_{p,u},\mathcal{R}_{p,u}\right)\big)\bigg), \label{e23a}\\
      \quad s.t. & \quad N_{T} \leq N_{p} + N_{d} \leq N_{tot}, \ \ N_{p} \geq N_{T}, \label{e23b}\\
                 & \quad \varepsilon_{c,u}\big(N_{d},\boldsymbol{\mathrm{\Gamma}}_{c,u},\mathcal{R}_{c}\big) \leq \varepsilon_{th}, \forall u \in \mathcal{U}, \label{e23c}\\
                 & \quad \varepsilon_{p,u}\big(N_{d},\boldsymbol{\mathrm{\Gamma}}_{p,u},\mathcal{R}_{p,u}\big) \leq \varepsilon_{th}, \forall u \in \mathcal{U}, \label{e23d}\\
                 & \quad 0 \leq \mathcal{R}_{c} \leq \min_{u \in \mathcal{U}}\big\{ C_{c,u}\!\left(N_{d},\boldsymbol{\mathrm{\Gamma}}_{\!c,u},\varepsilon_{th}\!\right)\!\big\},\label{e23e}\\
                 \vspace{-0.3em}
                 & \quad \sum_{u = 1}^{U} \mathcal{R}_{c,u} \leq \mathcal{R}_{c}\label{e23f},\\
                 & \quad \mathcal{R}_{min} \leq \mathcal{R}_{c,u} + \mathcal{R}_{p,u}, \forall u \in \mathcal{U}, \label{e23g}\\
                 & \quad \rho_{c} + \sum\limits_{u = 1}^{U}\rho_{p,u} \leq \rho_{tot}, \label{e23h}\\
                 \vspace{-0.3em}
                 % & \quad \rho_{p}N_{p} + N_{d}\big(\rho_{c} + \sum\limits_{u = 1}^{U}\rho_{p,u}\big) \leq E_{tot}, \label{e23i}\\
                 & \quad \mathcal{R}_{c,u} \geq 0,\ \mathcal{R}_{p,u} \geq 0, \forall u \in \mathcal{U}, \label{e23j}\\
                 & \quad \rho_{c} \geq 0,\ \rho_{p,u} \geq 0, \forall u \in \mathcal{U}, \label{e23k}
   \vspace{-1em}
   \end{align}
\end{subequations}
where $\mathcal{R}_{min}$ denotes the minimum transmission rate requirement for each receiver, and $\boldsymbol{\rho} \!=\! \left[\rho_{c},\rho_{p,1},\cdots,\rho_{p,U}\right]^{T}$ and $\boldsymbol{\mathcal{R}} = \left[\mathcal{R}_{c},\mathcal{R}_{p,1},\cdots,\mathcal{R}_{p,U}\right]^{T}$. (\ref{e23b}) delineates the latency constraints. (\ref{e23c}) and (\ref{e23d}) indicate the reliability constraints for receiver $u \in \mathcal{U}$ when delivering the common and private messages, respectively. (\ref{e23e}) and (\ref{e23f}) jointly define the constraints on common transmission rates; (\ref{e23g}) specifies the minimum transmission rate constraints for receivers, and (\ref{e23h}) ensures compliance with the total transmit power constraint.
\vspace{-0.7em}

\subsection{Problem Reformulation}

\vspace{-0.3em}
\par The solution of $\mathcal{P}1$ involves intricately intertwined joint optimization of $N_{p}$, $N_{d}$, $N_{T}$, $\boldsymbol{\mathcal{R}}$, and $\boldsymbol{\rho}$, leading to a non-convex objective function (\ref{e23a}). This poses $\mathcal{P}1$ becomes a formidable problem, rendering its optimal solution exceedingly challenging. For this reason, it is essential to excavate the intrinsic properties of $\mathcal{P}1$ to simplify and transform $\mathcal{P}1$. \textbf{Lemma 1} is elucidated to tackle this challenge.

\vspace{-0.4em}

\begin{lemma}\label{le1}
   Assuming receivers are uniformly distributed within the cell with an inner radius of $R_{min}$ and an outer radius of $R_{max}$. The BS is located at a height $\tilde{h}$, satisfying $\tilde{h} \ll R_{min}$ and $\tilde{h} \ll R_{max}$. Then, we can derive that
   \vspace{-0.5em}
  \begin{equation}\label{e24}
     \!\!\!\!\! \widetilde{\Pi}_{1} = \frac{\lambda^{2}\!\left(\!X_{max}^{(1)} \!-\! X_{min}^{(1)}\!\right)}{(4\pi)^{2}\!\left(R_{max}^{2}\!-\!R_{min}^{2}\right)},  \widetilde{\Pi}_{2} \!=\! \frac{\lambda^{2}\!\log\!\!\left(\frac{X_{max}^{(1)}}{X_{min}^{(1)}}\!\right)}{(4\pi)^{2}\!\left(R_{max}^{2}\!-\!R_{min}^{2}\right)},
  \end{equation}
  where
  \begin{equation*}
      \left\{
        \begin{array}{ll}
          \!\!\!\!\!\!\!\! & X_{min}^{(1)} = \frac{\lambda^{2}}{(4\pi)^{2}(R^{2}_{max}+\tilde{h}^{2})+N_{p}\rho_{p}\lambda^{2}},\\
          \!\!\!\!\!\!\!\! & X_{max}^{(1)} = \frac{\lambda^{2}}{(4\pi)^{2}(R^{2}_{min}+\tilde{h}^{2})+N_{p}\rho_{p}\lambda^{2}}.
        \end{array}
      \right.
  \end{equation*}

\vspace{-0.4em}

  \begin{equation}\label{e25}
    \!\!\!\!\!\!\!\!\!\!\!\!\! \widetilde{\Pi}_{3} \!=\! \left\{\!\!\!\!
                                                                                                                        \begin{array}{ll}
                                                                                                                          \frac{\lambda^{2}\log\left(\frac{R_{max}}{R_{min}}\right)}{8\left(\! \pi N_{p}\rho_{p}\!\right)^{2}\left(R_{max}^{2} - R_{min}^{2}\right)}, & \hbox{$N_{p}\rho_{p}\kappa_{u} \!\gg\! 1$;} \\
                                                                                                                          \frac{\lambda^{6}\!\left(R_{max}^{2} \!+\! R_{min}^{2}\right)}{2(4\pi)^{6}R_{max}^{4}R_{min}^{4}}, & \hbox{$0 < N_{p}\rho_{p}\kappa_{u} \!\ll\! 1$;} \\
                                                                                                                          \frac{\lambda^{6}\!\left(R_{max}^{2} \!+\! R_{min}^{2}\right)}{8(4\pi)^{6}R_{max}^{4}R_{min}^{4}}, & \hbox{$N_{p}\rho_{p}\kappa_{u} \!\approx\! 1$.}
                                                                                                                        \end{array}
                                                                                                                      \right.
  \end{equation}

  \vspace{-0.8em}

  \begin{equation}\label{e26}
    \widetilde{\Pi}_{4} = \left\{\!\!\!\!
             \begin{array}{ll}
               \frac{\lambda^{6}\left(R_{max}^{2} + R_{min}^{2}\right)}{(4\pi)^{6}\left(N_{p}\rho_{p}\right)^{3}R_{max}^{4}R_{min}^{4}}, \!\!\!\! & \hbox{$N_{p}\rho_{p}\kappa_{u} \!\gg\! 1$;} \\
               \frac{\lambda^{8}}{3(4\pi)^{8}\left(R_{max}^{2}-R_{min}^{2}\right)}\!\!\left(\!\!\frac{1}{R_{min}^{6}}-\frac{1}{R_{max}^{6}}\!\!\right), \!\!\!\! & \hbox{$0 < N_{p}\rho_{p}\kappa_{u} \!\ll\! 1$;} \\
               \frac{\lambda^{8}}{12(4\pi)^{8}\left(R_{max}^{2}-R_{min}^{2}\right)}\left(\frac{1}{R_{min}^{6}}-\frac{1}{R_{max}^{6}}\right), \!\!\!\! & \hbox{$N_{p}\rho_{p}\kappa_{u} \!\approx\! 1$.}
             \end{array}
           \right.
  \end{equation}

  \vspace{-1em}

  \begin{equation}\label{e27}
   \!\!\!\!\!\!\!\!\!\!\!\!\!\!\!\!\!\!\!\!\!\!\!\!\!\! \widetilde{\Pi}_{5} \!=\! \left\{\!\!\!\!
              \begin{array}{ll}
                \frac{\lambda^{2}\log\left(\frac{R_{max}}{R_{min}}\right)}{8\pi^{2} N_{p}\rho_{p} \left(R_{max}^{2} - R_{min}^{2}\right)}, & \hbox{$N_{p}\rho_{p}\kappa_{u} \!\gg\! 1$;}\\
                \frac{\lambda^{6}\left(R_{max}^{2} + R_{min}^{2}\right)}{(4\pi)^{6}\left(N_{p}\rho_{p}\right)^{3}R_{max}^{4}R_{min}^{4}}, & \hbox{$0 \!<\! N_{p}\rho_{p}\kappa_{u} \!\ll\! 1$;}\\
                \frac{\lambda^{6}\left(R_{max}^{2} + R_{min}^{2}\right)}{2(4\pi)^{6}\left(N_{p}\rho_{p}\right)^{3}R_{max}^{4}R_{min}^{4}},& \hbox{$N_{p}\rho_{p}\kappa_{u} \!\approx\! 1$;}
              \end{array}
            \right.
  \end{equation}
  where \\
  $\small{\widetilde{\Pi}_{1} \!=\! \mathbb{E}\!\!\left[\!\frac{\kappa_{u}^{2}}{\left(1 + N_{p}\rho_{p}\kappa_{u}\right)^{2}}\!\right]}$,$\small{\widetilde{\Pi}_{2} \!=\! \mathbb{E}\!\!\left[\!\frac{\kappa_{u}^{3/2}}{1 + N_{p}\rho_{p}\kappa_{u}}\!\right]}$,$\small{\widetilde{\Pi}_{3} \!=\! \mathbb{E}\!\!\left[\!\frac{\kappa_{u}^{3}}{\left(1+N_{p}\rho_{p}\kappa_{u}\right)^{2}}\!\right]}$,
  \begin{small}
  \begin{equation*}
  \!\!\!\!\!\!\!\!\!\!\!\!\!\!\!\!\!\!\!\!\!\!\!\!\!\!\!\!\!\!\!\!\! \widetilde{\Pi}_{4} = \mathbb{E}\!\!\left[\frac{\kappa_{u}^{4}}{\left(1 + N_{p}\rho_{p}\kappa_{u}\right)^{2}}\right], \text{\rm{and} } \widetilde{\Pi}_{5} = \mathbb{E}\!\!\left[\!\frac{\kappa_{u}^{2}}{1 \!+\! N_{p}\rho_{p}\kappa_{u}}\!\right].
  \end{equation*}
  \end{small}
\end{lemma}

\vspace{-1.0em}

\begin{proof}
  The proof of \textbf{Lemma 1} is given in Appendix A.
\end{proof}

\vspace{-0.7em}

Using \textbf{Lemma 1}, \textbf{Theorem 2} is further derived as follows:

\vspace{-0.5em}

\begin{theorem}\label{theo1}
   Under the proposed zero-forcing precoding scheme, the closed-form expressions for the SINR of common and private messages corresponding to (\ref{e14}), (\ref{e15}), and (\ref{e16}) can be transformed into the following forms:
   \begin{subequations}\label{e28}
      \begin{align}
       & \Gamma_{c,u}^{(j)} = \frac{\rho_{c}\kappa_{u}\Psi}{\kappa_{u}\Psi\sum_{k=1}^{U}\rho_{p,k} + 1} \!\approx\! \frac{\rho_{c}}{\sum\limits_{k=1}^{U}\!\!\rho_{p,k}},\\
       & \Gamma_{p,u}^{(j)} = \frac{\rho_{p,u}\kappa_{u}\Psi}{\kappa_{u}\Psi\sum\limits_{k=1,k \neq u}^{U}\rho_{p,k} + 1} \!\approx\! \frac{\rho_{p,u}}{\sum\limits_{k=1,k \neq u}^{U}\!\!\!\rho_{p,k}}, \\
       & \widehat{\Gamma}_{p,u}^{(j)} = \frac{\rho_{p,u}\kappa_{u}\Psi}{\kappa_{u}\Psi \sum\limits_{k=1,k \neq u}^{U}\rho_{p,k} + \rho_{c}\kappa_{u}\Psi + 1} \!\approx\! \frac{\rho_{p,u}}{\sum\limits_{k=1,k\neq u}^{U}\!\!\!\rho_{p,k} \!+\! \rho_{c}},
       \vspace{-0.5em}
      \end{align}
   \end{subequations}
   \vspace{-0.5em}
    where the expression of $\Psi$ can be given by (\ref{e36}).
\end{theorem}

\vspace{-1em}

\begin{proof}
   According to (\ref{e14}), we can expand $\mathrm{Var}\left[\sqrt{\rho_{c}} \boldsymbol{\mathrm{g}}_{u}^{(j)} \boldsymbol{\mathrm{w}}_{c} \boldsymbol{\mathrm{s}}_{c}\right]$ into the form described in (\ref{e30}), where $(a)$ can be directly obtained by substituting (\ref{e10}), $(b)$ is easily derived by substituting (\ref{e6}), and $(c)$ is obtained from the property of independent random variables $X$ and $Y$, namely, $\mathbb{E}\big[\left|X+Y\right|^{2}\big] = \mathbb{E}\big[\left|X\right|^{2}\big]+\mathbb{E}\big[\left|Y\right|^{2}\big]$. Subsequently, we further expand the two components of $\mathrm{Var}\left[\sqrt{\rho_{c}} \boldsymbol{\mathrm{g}}_{u}^{(j)} \boldsymbol{\mathrm{w}}_{c} \boldsymbol{\mathrm{s}}_{c}\right]$ in (\ref{e30}), as formulated in (\ref{e31}) and (\ref{e32}), respectively. $(d)$ and $(f)$ are expanded through the distributive law of multiplication, followed by organizing and combining the resulting expressions. $(e)$ and $(g)$ can be derived from (\ref{e33}) and (\ref{e34}), respectively, as follows:

\vspace{-1em}

\begin{figure*}[t]
\centering
\hrulefill
\begin{equation}\label{e30}
\setstretch{0.95}
  \begin{aligned}
  & \mathrm{Var}\left[\sqrt{\rho_{c}} \boldsymbol{\mathrm{g}}_{u}^{(j)} \boldsymbol{\mathrm{w}}_{c} \boldsymbol{\mathrm{s}}_{c}\right] = \rho_{c}\mathbb{E}\left[\left|\boldsymbol{\mathrm{g}}_{u}^{(j)} \boldsymbol{\mathrm{w}}_{c}\right|^{2}\right] = \rho_{c}\mathbb{E}\bigg[\bigg(\sum\limits_{n = 1}^{N_{T}}g_{u,n}^{(j)}w_{c,n}\bigg)^{2}\bigg] \overset{(a)}{=} \rho_{c}\mathbb{E}\bigg[\bigg(\sum\limits_{n = 1}^{N_{T}}g_{u,n}^{(j)} \sum\limits_{k = 1}^{U} \xi_{k} \widehat{g}_{k,n}^{(j)}\bigg)^{2}\bigg] \overset{(b)}{=} \rho_{c}\mathbb{E}\bigg[\bigg(\sum\limits_{n = 1}^{N_{T}} \sum\limits_{k = 1}^{U} g_{u,n}^{(j)}\xi_{k} \cdot \\
 & \bigg[\!\frac{N_{p}\rho_{p}\kappa_{k}}{1 + N_{p}\rho_{p}\kappa_{k}}g_{k,n}^{(j)} + \frac{\sqrt{N_{p}\rho_{p}}\kappa_{k}}{1 \!+\! N_{p}\rho_{p}\kappa_{k}}n_{p}^{(j)}\bigg]\bigg)^{2}\bigg] \overset{(c)}{=} \rho_{c} \mathbb{E}\bigg[\!\bigg(\!\sum\limits_{n = 1}^{N_{T}} \sum\limits_{k = 1}^{U} g_{u,n}^{(j)} \underbrace{\frac{N_{p}\rho_{p}\kappa_{k}\xi_{k}}{1 \!+\! N_{p}\rho_{p} \kappa_{k}}g_{k,n}^{(j)}}_{D_{k,n}}\!\bigg)^{\!2}\bigg] + \rho_{c}\mathbb{E}\bigg[\!\bigg(\!\sum\limits_{n = 1}^{N_{T}} \sum\limits_{k = 1}^{U} g_{u,n}^{(j)} \underbrace{\frac{\sqrt{N_{p}\rho_{p}}\kappa_{k}\xi_{k}}{1 + N_{p}\rho_{p}\kappa_{k}}}_{F_{k}}n_{p}^{(j)}\!\bigg)^{\!2}\bigg]
  \end{aligned}
\end{equation}
\vspace{-0.2em}
\hrulefill
\vspace{-0.5em}
\begin{small}
\begin{equation}\label{e31}
  \begin{aligned}
     & \rho_{c}\mathbb{E}\!\bigg[\!\bigg(\!\sum\limits_{n = 1}^{N_{T}} \! \sum\limits_{k = 1}^{U} g_{u,n}^{(j)} F_{k}n_{p}^{(j)}\!\bigg)^{\!2}\bigg] \! \overset{(d)}{=} \! \rho_{c} \! \sum\limits_{n = 1}^{N_{T}} \! \sum\limits_{k = 1}^{U} \mathbb{E}\! \bigg[\!\big(g_{u,n}^{(j)} F_{k}\big)^{2}\bigg] \!+\! \rho_{c}\mathbb{E}\!\bigg[\sum\limits_{n = 1}^{N_{T}}\!\left(g_{u,n}^{(j)}\right)^{2}\sum\limits_{k = 1}^{U}\!\!\bigg(\!\!F_{k}\!\!\!\!\sum\limits_{l = 1, l \neq k}^{U}\!\!\!\!F_{l}\!\bigg)\!\bigg] \!+\! \rho_{c}\mathbb{E}\bigg[\sum\limits_{n = 1}^{N_{T}} \sum\limits_{k = 1}^{U} g_{u,n}^{(j)} F_{k}\!\bigg(\!\sum\limits_{t = 1, t \neq n}^{N_{T}}\sum\limits_{l = 1}^{U}\!\! g_{u,t}^{(j)}F_{l}\!\bigg)\!\bigg]\\
    & \overset{(e)}{=} \rho_{c}N_{T}\kappa_{u}\bar{h}^{2}N_{p}\rho_{p}\mathbb{E}\!\!\left[\!\frac{\kappa_{u}^{2}}{\left(1 + N_{p}\rho_{p}\kappa_{u}\right)^{2}}\!\right]\sum\limits_{k=1}^{U}\xi_{k}^{2} + \rho_{c}N_{T}\kappa_{u}\bar{h}^{2}N_{p}\rho_{p}\left(\mathbb{E}\!\!\left[\!\frac{\kappa_{u}}{1 + N_{p}\rho_{p}\kappa_{u}}\!\right]\right)^{2}\sum\limits_{k=1}^{U}\sum\limits_{v=1,u \neq k}^{U}\xi_{v}\xi_{v} + \rho_{c}N_{T}(N_{T}-1)\kappa_{u}\bar{h}^{2}N_{p}\rho_{p}\times \\
    & \ \ \ \ \ \left(\mathbb{E}\!\!\left[\!\frac{\kappa_{u}^{2}}{\left(1 + N_{p}\rho_{p}\kappa_{u}\right)^{2}}\!\right]\sum\limits_{k=1}^{U}\xi_{k}^{2} + \left(\mathbb{E}\!\!\left[\!\frac{\kappa_{u}}{1 + N_{p}\rho_{p}\kappa_{u}}\!\right]\right)^{2}\sum\limits_{k=1}^{U}\sum\limits_{v=1,u \neq k}^{U}\xi_{v}\xi_{v}\right).
  \end{aligned}
\end{equation}
\end{small}
\vspace{-0.2em}
\hrulefill
\vspace{-0.5em}
\begin{small}
\begin{equation}\label{e32}
   \begin{aligned}
     & \rho_{c} \mathbb{E}\bigg[\!\bigg(\!\sum\limits_{n = 1}^{N_{T}} \sum\limits_{k = 1}^{U} g_{u,n}^{(j)} D_{k,n}\!\!\bigg)^{\!2}\bigg] \overset{(f)}{=} \rho_{c}\sum\limits_{n = 1}^{N_{T}}\sum\limits_{k = 1}^{U} \mathbb{E}\bigg[\!\!\left(g_{u,n}^{(j)}D_{k,n}\!\right)^{2}\!\bigg] \!+\! \rho_{c}\mathbb{E}\bigg[\sum\limits_{n = 1}^{N_{T}}\left(g_{u,n}^{(j)}\right)^{2}\sum\limits_{k = 1}^{U}\bigg(\!D_{k,n}\!\!\!\!\!\sum\limits_{v = 1, v \neq k}^{U}\!\!\!\!\!D_{v,n}\!\!\bigg)\bigg] + \rho_{c}\mathbb{E}\bigg[\sum\limits_{n = 1}^{N_{T}}\sum\limits_{k = 1}^{U}g_{u,n}^{(j)}D_{k,n}\bigg(\sum\limits_{t = 1, t\neq n}^{N_{T}}\\
     & \ \ \ \sum\limits_{v = 1}^{U}g_{u,t}^{(j)}D_{v,n}\bigg)\bigg] \overset{(g)}{=} \rho_{c}N_{T}\kappa_{u}\bar{h}^{4}\left(N_{p}\rho_{p}\right)^{2} \mathbb{E}\!\!\left[\!\frac{\kappa_{u}^{3}}{\left(1+N_{p}\rho_{p}\kappa_{u}\right)^{2}}\!\right]\sum\limits_{k=1}^{U}\xi_{k}^{2} + \rho_{c}N_{T}\kappa_{u}\bar{h}^{4}\left(N_{p}\rho_{p}\right)^{2}\bigg(\mathbb{E}\!\!\left[\!\frac{\kappa_{u}^{3/2}}{1 + N_{p}\rho_{p}\kappa_{u}}\!\right]\bigg)^{2} \sum\limits_{k = 1}^{U}\sum\limits_{v=1,v\neq k}\xi_{k}\xi_{v} \\
    & \ \ \ + \rho_{c}N_{T}\left(N_{T}-1\right)\kappa_{u}\bar{h}^{4}\left(N_{p}\rho_{p}\right)^{2}\bigg(\mathbb{E}\!\!\left[\frac{\kappa_{u}^{4}}{\left(1 + N_{p}\rho_{p}\kappa_{u}\right)^{2}}\right]\sum\limits_{k=1}^{U}\xi_{k}^{2}+\left(\mathbb{E}\!\!\left[\!\frac{\kappa_{u}^{2}}{1 \!+\! N_{p}\rho_{p}\kappa_{u}}\!\right]\right)^{2}\sum\limits_{k=1}^{U}\sum\limits_{v=1,v\neq k}\xi_{k}\xi_{v}\bigg).
   \end{aligned}
\end{equation}
\end{small}
\vspace{-0.2em}
\hrulefill
\vspace{-1.5em}
\end{figure*}

\begin{subequations}\label{e33}
   \begin{align}
     \mathbb{E}\big[F_{k}^{2}\big] & =  N_{p}\rho_{p}\xi_{k}\bar{h}^{2} \mathbb{E}\!\!\left[\!\frac{\kappa_{u}^{2}}{\left(1 + N_{p}\rho_{p}\kappa_{u}\right)^{2}}\!\right],\\
     \mathbb{E}\big[F_{k}\big] & = \sqrt{N_{p}\rho_{p}}\xi_{k}\bar{h} \mathbb{E}\!\!\left[\!\frac{\kappa_{u}}{\left(1 + N_{p}\rho_{p}\kappa_{u}\right)}\!\right],
   \end{align}
\end{subequations}

\vspace{-0.8em}

\begin{subequations}\label{e34}
   \begin{align}
    \quad \quad \quad \ \ \mathbb{E}\big[D_{k,n}^{2}\big] & =  \big(N_{p}\rho_{p}\xi_{k}\bar{h}\big)^{2}\mathbb{E}\!\!\left[\!\frac{\kappa_{u}^{3}}{\left(1 + N_{p}\rho_{p}\kappa_{u}\right)^{2}}\!\right], \\
    \quad \quad \quad \ \ \mathbb{E}\big[D_{k,n}\big] & = N_{p}\rho_{p}\xi_{k}\bar{h}\mathbb{E}\!\!\left[\!\frac{\kappa_{u}^{3/2}}{1 + N_{p}\rho_{p}\kappa_{u}}\!\right].
   \end{align}
\end{subequations}

\par By substituting (\ref{e31}) and (\ref{e32}) into (\ref{e30}) and rearranging (\ref{e30}) leveraging \textbf{Lemma 1}, we can derive that
\begin{equation}\label{e35}
   \mathrm{Var}\left[\sqrt{\rho_{c}} \boldsymbol{\mathrm{g}}_{u}^{(j)} \boldsymbol{\mathrm{w}}_{c} \boldsymbol{\mathrm{s}}_{c}\right] = \rho_{c}\kappa_{u}\Psi,
\end{equation}
where $\Psi$ is given by
\vspace{-0.6em}
\begin{small}
\begin{equation}\label{e36}
   \begin{aligned}
   \Psi & \!=\! N_{T}\bar{h}^{2}N_{p}\rho_{p}\bigg[\!\big(\!\sum\limits_{k=1}^{U}\xi_{k}^{2}\big)\!\!\left[\!\bar{h}^{2}N_{p}\rho_{p}\!\left(\widetilde{\Pi}_{3}\!+\!(\!N_{T}\!-\!1)\widetilde{\Pi}_{4}\!\right) \!+\! N_{T}\widetilde{\Pi}_{2}\right] \\
   & + \!\big(\sum\limits_{k=1}^{U}\sum\limits_{v=1,v\neq u}^{U}\!\!\! \xi_{k}\xi_{v}\! \big)\!\big[\bar{h}^{2}N_{p}\rho_{p}\!\left(\widetilde{\Pi}_{1}^{2} \!+\!\left(N_{T}\!-\!1\right)\widetilde{\Pi}_{5}^{2}\right) \!+\! N_{T}\widetilde{\Pi}_{1}^{2}\big]\bigg].
   \end{aligned}
\end{equation}
\end{small}

\vspace{-1.2em}

\par Similarly, we can derive that
\vspace{-0.6em}
\begin{equation}\label{e37}
   \sum_{k = 1}^{U}\mathrm{Var}\left[\sqrt{\rho_{p,k}} \boldsymbol{\mathrm{g}}_{u}^{(j)} \boldsymbol{\mathrm{w}}_{p,k} \boldsymbol{\mathrm{s}}_{p,k}\right] = \kappa_{u}\Psi \sum\limits_{k=1}^{U}\rho_{p,k}.
\end{equation}

\par By substituting (\ref{e36}) and (\ref{e37}) into (\ref{e14}), we can derive (\ref{e28}a). Similarly, (\ref{e28}b) and (\ref{e28}c) can also be derived.
\end{proof}

\vspace{-0.8em}

\par \textbf{Theorem 2} uncovers the intrinsic interrelationship between SINRs and the transmit power of the decoded and interfering streams, holding the potential to transform complex matrix optimization problems into constant-variable optimization problems. Leveraging \textbf{Theorem 2}, $\mathcal{P}1$ can be transformed into $\mathcal{P}2$ as follows:

\vspace{-1.5em}

\begin{subequations}
   \begin{align}
      \!\!\!\!\!\!\!\!\!\!\!\!\! \mathcal{P}2:&\ \max_{\{N_{p},N_{d},N_{T},\boldsymbol{\mathcal{R}},\boldsymbol{\rho}\}} \!\! \mathcal{T}\big(N_{p},N_{d},N_{T},\boldsymbol{\mathcal{R}},\boldsymbol{\rho}\big) \!=\! \sum\limits_{u \in \mathcal{U}} \!\! \bigg(\! \mathcal{R}_{c,u} \! \cdot \! \big(1 - \nonumber \\
      & \varepsilon_{c,u}\!\big(N_{d},\boldsymbol{\rho},\mathcal{R}_{c}\!\big)\!\big) \!+\! \mathcal{R}_{p,u}\!\cdot\!\big(1 \!-\! \varepsilon_{p,u}\!\left(N_{d},\boldsymbol{\rho},\mathcal{R}_{p,u}\right)\!\big)\!\!\bigg), \label{e38a}\\
      \quad s.t. & \quad \varepsilon_{c,u}\big(N_{d},\boldsymbol{\rho},\mathcal{R}_{c}\big) \leq \varepsilon_{th}, \forall u \in \mathcal{U}, \label{e38b}\\
                 & \quad \varepsilon_{p,u}\big(N_{d},\boldsymbol{\rho},\mathcal{R}_{p,u}\big) \leq \varepsilon_{th}, \forall u \in \mathcal{U}, \label{e38c}\\
                 & \quad 0 \leq \mathcal{R}_{c} \leq \min_{u \in \mathcal{U}}\big\{ C_{c,u}\!\left(N_{d},\boldsymbol{\rho},\varepsilon_{th}\!\right)\!\big\},\label{e38d}\\
                 \vspace{-0.3em}
                 & \quad (\ref{e23b}), \mathrm{and}\ (\ref{e23f})-(\ref{e23k}) \label{e38e}.
   \vspace{-1em}
   \end{align}
\end{subequations}

\vspace{-0.3em}

\par \par After reformulating $\mathcal{P}1$, $\mathcal{P}2$ with a more concise form is derived. Nevertheless, tackling $\mathcal{P}2$ remains highly challenging, primarily due to the non-convex nature of $Q\big(g\left(N_{d},\boldsymbol{\rho},\varepsilon\right)\big)$ and the reliability constraints $(\ref{e38b})$-$((\ref{e38c})$. To effectively address $\mathcal{P}2$, an optimality analysis for $\mathcal{P}2$ is conducted.

\vspace{-0.6em}

\subsection{Optimality Analysis}
\vspace{-0.3em}
\par Observing $\mathcal{P}2$ intuitively reveals two key points. Firstly, increasing the allocated split rate deteriorates receivers' DEP performance, potentially violating reliability constraints. Secondly, allocating more transmit power and CUs to uplink pilot training improves the channel estimation quality (i.e., $\bar{w}_{u} \!\rightarrow \! 0$), however, it reduces the information bits available for downlink data transmission. Therefore, it is essential to judiciously allocate the split rate, blocklength, and transmit power to optimally balance the resource overhead between uplink pilot training and downlink short-packet transmissions. Inspired by the above analysis, \textbf{Lemma 2} is presented below:

\vspace{-0.4em}

\begin{lemma}\label{le2}
   The DEP $\varepsilon_{c,u}\big(N_{d},\boldsymbol{\rho},\mathcal{R}_{c}\big)$$/$$\varepsilon_{p,u}\big(N_{d},\boldsymbol{\rho},\mathcal{R}_{p,u}\big)$ strictly decreases with $N_{d}$ and $\rho_{c}$$/$$\rho_{p,u}$, and strictly increases with $\mathcal{R}_{c}$$/$$\mathcal{R}_{p,u}$.
\end{lemma}

\vspace{-1em}

\begin{proof}
   The proof of \textbf{Lemma 2} is given in Appendix B.
\end{proof}

\vspace{-0.6em}

\par \textbf{Corollary 1} is derived to reveal the optimal conditions. % for $\mathcal{P}2$.

\vspace{-0.6em}

\begin{corollary}\label{coro1}
   The reliability constraints (\ref{e38b}) and (\ref{e38c}) are active at the optimal solutions. Moreover, there exists at least one optimal solution such that $N_{p} = N_{T}$ holds.
\end{corollary}

\vspace{-1.2em}

\begin{proof}
  The proof of \textbf{Corollary 1} is given in Appendix C.
\end{proof}

\vspace{-0.6em}

\par According to \textbf{Lemma 2} and \textbf{Corollary 1}, $\mathcal{P}2$ can be equivalently restated as

\vspace{-1.2em}

\begin{subequations}
   \begin{align}
      \widetilde{\mathcal{P}}2:&\ \max_{\{N_{T},\boldsymbol{\mathcal{R}},\boldsymbol{\rho}\}} \mathcal{T}\big(N_{T},\boldsymbol{\mathcal{R}},\boldsymbol{\rho}\big), \label{e39a}\\
      \quad s.t. & \quad \sum_{u=1}^{U}\mathcal{R}_{c,u} = \mathcal{R}_{c},\label{e39b}\\
                 \vspace{-0.8em}
                 % & \quad \rho_{p}N_{T} + N_{d}\big(\rho_{c}+\sum_{u=1}^{U}\rho_{p,u}\big) = E, \label{e39c}\\
                 \vspace{-0.5em}
                 & \quad (\ref{e38b})-(\ref{e38d}),(\ref{e23g}), (\ref{e23h}), (\ref{e23j}), (\ref{e23k}). \label{e39c}
   \vspace{-1em}
   \end{align}
\end{subequations}

\vspace{-0.6em}

\par Although $\widetilde{\mathcal{P}}2$ has been simplified, it remains a highly coupled non-convex optimization problem. In this case, $\widetilde{\mathcal{P}}2$ is decoupled into three subproblems: the first one involves power allocation, the second one focuses on rate-splitting optimization, furthermore, building upon the results of \textbf{Subproblem I} and \textbf{II}, the third one deals with transmit antenna selection.

\vspace{-0.8em}

\subsection{Subproblem I: Power Allocation Optimization}
\vspace{-0.2em}
\par Given the rate-splitting strategy $\overline{\boldsymbol{\mathcal{R}}}$, $\widetilde{\mathcal{P}}2$ depends on $\boldsymbol{\rho}$ and $N_{T}$. Temporarily disregarding $N_{T}$, \textbf{Subproblem I} can be formulated as follows:
\begin{subequations}\label{e40}
   \begin{align}
      \widetilde{\mathcal{P}}2\mathrm{-I}:&\ \ \max_{\{\boldsymbol{\rho}\}} \mathcal{T}\big(\overline{\boldsymbol{\mathcal{R}}},\boldsymbol{\rho}\big), \label{e40a}\\
      \quad s.t. & \quad (\ref{e38b}), (\ref{e38c}), (\ref{e39c}), (\ref{e23k}). \label{e40b}
   \vspace{-1em}
   \end{align}
\end{subequations}

\vspace{-0.5em}

\begin{lemma}\label{le3}
   The function $g(N_{d},\boldsymbol{\Gamma},R)$ is a monotonically increasing concave function with respect to $\boldsymbol{\Gamma}$.
\end{lemma}

\vspace{-1.2em}

\begin{proof}
   The proof of \textbf{Lemma 3} is given in Appendix D.
\end{proof}

\vspace{-0.7em}

\par $\widetilde{\mathcal{P}}2\mathrm{-I}$ is a non-convex problem since (\ref{e40a}), (\ref{e38b}), and (\ref{e38c}) are non-convex. To effectively tackle this intractable non-convex subproblem, we leverage the SCA technique to handle (\ref{e40a}), (\ref{e38b}), and (\ref{e38c}). Firstly, we introduce auxiliary variables $\boldsymbol{\mathcal{A}} \triangleq \left[\mathscr{A}_{1},\cdots,\mathscr{A}_{U}\right]^{T}$ and $\boldsymbol{\mathcal{B}} \triangleq \left[\mathscr{B}_{1},\cdots,\mathscr{B}_{U}\right]^{T}$, then $\widetilde{\mathcal{P}}2\mathrm{-I}$ can be equivalently transformed into $\mathcal{P}3$, as follows:

\vspace{-1.5em}

\begin{subequations}\label{e41}
   \begin{align}
      \widetilde{\mathcal{P}}3:&\ \max_{\{\boldsymbol{\mathcal{A}},\boldsymbol{\mathcal{B}},\boldsymbol{\rho}\}}\sum\limits_{u=1}^{U}\!\!\bigg(\!\mathcal{R}_{c,u}\big(1 \!-\!Q(\mathscr{A}_{u})\big) \!+\! \mathcal{R}_{p,u}\big(1 \!-\! Q(\mathscr{B}_{u})\big)\!\!\bigg), \label{e41a}\\
      s.t. &\ Q(\mathscr{A}_{u}) \leq \varepsilon_{th}, Q(\mathscr{B}_{u}) \leq \varepsilon_{th}, \forall u \in \mathcal{U}, \label{e41b}\\
                 \vspace{-0.5em}
           &\ \mathscr{A}_{u} \leq \ln 2 \bigg(\!\sum\limits_{j=1}^{N_{R}}\log_{2}\!\big(1\!+\!\Gamma_{c,u}^{(j)}\big)\!-\!\overline{\mathcal{R}}_{c,u}\!\!\bigg)\!\!\cdot\!\!\bigg(\!\sum\limits_{j=1}^{N_{R}}\!\!\sqrt{\frac{\mathcal{V}(\Gamma_{c,u}^{(j)})}{N_{d}}}\bigg)^{-1}\!\!\!, \label{e41c}\\
           &\ \mathscr{B}_{u} \leq \ln 2 \bigg(\!\sum\limits_{j=1}^{N_{R}}\log_{2}\!\big(1\!+\!\Gamma_{p,u}^{(j)}\big)\!-\!\overline{\mathcal{R}}_{p,u}\!\!\bigg)\!\!\cdot\!\!\bigg(\!\sum\limits_{j=1}^{N_{R}}\!\!\sqrt{\frac{\mathcal{V}(\Gamma_{p,u}^{(j)})}{N_{d}}}\bigg)^{-1}\!\!\!, \label{e41c}\\
                 \vspace{-0.3em}
           &\ (\ref{e39c}), \mathrm{and}\ (\ref{e23k}), \label{e41d}
   \vspace{-1em}
   \end{align}
\end{subequations}
where (\ref{e41a}) is a concave function and (\ref{e41b}) is a convex constraint, as $Q(x)$ is convex for $x > 0$. Nevertheless, (\ref{e41c}) and (\ref{e41d}) are non-convex constraints. % Remarkably, the function $g(N_{d},\boldsymbol{\Gamma},R)$ in (\ref{e2}) is strictly increasing and concave with respect to $\boldsymbol{\Gamma}$, which can be derived from \textbf{Lemma 3}.
Leveraging \textbf{Theorem 1} and combining \textbf{Lemma 3}, we can further introduce auxiliary variables $\boldsymbol{\mathcal{C}} \triangleq \left[\mathscr{C}_{1},\cdots,\mathscr{C}_{U}\right]^{T}$ and $\boldsymbol{\mathcal{D}} \triangleq \left[\mathscr{D}_{1},\cdots,\mathscr{D}_{U}\right]^{T}$, and then we have

\vspace{-1.5em}

\begin{small}
\begin{equation}\label{e42}
    \mathscr{A}_{u} \leq \frac{\sqrt{N_{d}}\ln 2}{N_{R}\sqrt{\mathcal{V}(\mathscr{C}_{u})}} \bigg(\!N_{R}\log_{2}\!\big(1\!+\!\mathscr{C}_{u}\big)\!-\!\overline{\mathcal{R}}_{c,u}\!\!\bigg), \forall u \in \mathcal{U},
\end{equation}
\end{small}

\vspace{-1em}

\begin{small}
\begin{equation}\label{e43}
    \mathscr{B}_{u} \leq \frac{\sqrt{N_{d}}\ln 2}{N_{R}\sqrt{\mathcal{V}(\mathscr{D}_{u})}} \bigg(\!N_{R}\log_{2}\!\big(1\!+\!\mathscr{D}_{u}\big)\!-\!\overline{\mathcal{R}}_{p,u}\!\!\bigg), \forall u \in \mathcal{U},
\end{equation}
\end{small}
\vspace{-0.5em}
where
\begin{equation}\label{e44}
   \!\!\!\!\! \mathscr{C}_{u} \! \leq \! \Gamma_{c,u}^{(j)} \! \approx \! \frac{\rho_{c}}{\sum_{k=1}^{U}\rho_{p,k}},\ \mathscr{D}_{u} \! \leq \! \Gamma_{p,u}^{(j)} \! \approx \! \frac{\rho_{p,u}}{\sum_{k=1, k \neq u}^{U}\rho_{p,k}}.
\end{equation}

\vspace{-0.8em}

\par Constraints (\ref{e42}) and (\ref{e43}) are both convex, however, constraint (\ref{e44}) is non-convex. For this reason, auxiliary variables $\boldsymbol{\mathcal{E}} \triangleq \left[\mathscr{E}_{1},\cdots,\mathscr{E}_{U}\right]^{T}$ and $\boldsymbol{\mathcal{F}} \triangleq \left[\mathscr{F}_{1},\cdots,\mathscr{F}_{U}\right]^{T}$ are introduced:

\vspace{-1.5em}

\begin{subequations}\label{e45}
   \begin{align}
      & \mathscr{C}_{u} \leq \frac{2\rho_{c}}{\mathscr{E}_{u}^{[t]}} - \frac{\rho_{c}\mathscr{E}_{u}}{\big(\mathscr{E}_{u}^{[t]}\big)^{2}}, \label{e45a}\\
      & \mathscr{E}_{u} \geq \sum\limits_{k=1}^{U}\rho_{p,k}, \label{e45b}\\
      & \mathscr{D}_{u} \leq \frac{2\rho_{p,u}}{\mathscr{F}_{u}^{[t]}} - \frac{\rho_{p,u}\mathscr{F}_{u}}{\big(\mathscr{F}_{u}^{[t]}\big)^{2}}, \label{e45c}\\
      & \mathscr{F}_{u} \geq \sum\limits_{k=1,k \neq u}^{U}\rho_{p,k}, \label{e45d}
   \end{align}
\end{subequations}
where constraints (\ref{e45a})$-$(\ref{e45d}) are all convex, and operator $[t]$ indicates the $t$-th iteration of the SCA method.

\par Combining (\ref{e40})-(\ref{e45}), $\widetilde{\mathcal{P}}2\mathrm{-I}$ can be finally equivalently reformulated as follows:

\vspace{-1.2em}

\begin{subequations}\label{e46}
   \begin{align}
     \widetilde{\mathcal{P}}4: \!\!\! &\ \max_{\{\boldsymbol{\mathcal{A}},\boldsymbol{\mathcal{B}},\boldsymbol{\mathcal{C}},\boldsymbol{\mathcal{D}},\boldsymbol{\mathcal{E}},\boldsymbol{\mathcal{F}},\boldsymbol{\rho}\}}\sum\limits_{u=1}^{U}\!\!\bigg(\!\!\mathcal{R}_{c,u}\!\big(1 \!-\!Q(\!\mathscr{A}_{u}\!)\big) \!+\! \mathcal{R}_{p,u}\!\big(1 \!-\! Q(\!\mathscr{B}_{u}\!)\big)\!\!\bigg), \label{e46a}\\
      s.t. &\ (\ref{e42}), (\ref{e43}), (\ref{e45a})-(\ref{e45d}), (\ref{e39c}), \mathrm{and}\ (\ref{e23k}).
   \end{align}
\end{subequations}

\vspace{-0.5em}

\par $\widetilde{\mathcal{P}}4$ is convex and can be efficiently solved using interior-point methods. The detailed solution process for \textbf{Subproblem I} is summarized as \textbf{Algorithm 1}.

\vspace{-0.7em}

\begin{algorithm}[h]% [htbp]
\small
\setstretch{0.7}
        \caption{SCA-based Algorithm for Solving Subproblem I.}
        \KwIn{$\overline{\boldsymbol{\mathcal{R}}} \triangleq \{\overline{\mathcal{R}}_{c},\overline{\mathcal{R}}_{c,1},\cdots,\overline{\mathcal{R}}_{c,U}\}$ and $N_{d}$;
}
        \textbf{Initialize:} $\boldsymbol{\mathcal{A}}^{[0]}$, $\boldsymbol{\mathcal{B}}^{[0]}$, $\boldsymbol{\mathcal{C}}^{[0]}$, $\boldsymbol{\mathcal{D}}^{[0]}$, $\boldsymbol{\mathcal{E}}^{[0]}$, $\boldsymbol{\mathcal{F}}^{[0]}$, $\boldsymbol{\rho}^{[0]}$;\\
        Set iteration index $t = 0$;\\
        \While{\text{Convergence Unsatisfied}}{
           Use $\boldsymbol{\mathcal{A}}^{[t]}, \boldsymbol{\mathcal{B}}^{[t]}, \boldsymbol{\mathcal{C}}^{[t]}, \boldsymbol{\mathcal{D}}^{[t]}, \boldsymbol{\mathcal{E}}^{[t]}, \boldsymbol{\mathcal{F}}^{[t]}, \boldsymbol{\rho}^{[t]}$ to solve convex problem $\widetilde{\mathcal{P}}4$ ;\\
           Obtain the optimal solution of $\widetilde{\mathcal{P}}4$: $\boldsymbol{\mathcal{A}}^{[t+1]}, \boldsymbol{\mathcal{B}}^{[t+1]}, \boldsymbol{\mathcal{C}}^{[t+1]}, \boldsymbol{\mathcal{D}}^{[t+1]}, \boldsymbol{\mathcal{E}}^{[t+1]}, \boldsymbol{\mathcal{F}}^{[t+1]}, \boldsymbol{\rho}^{[t+1]}$;\\
           Set iteration index $t = t+1$;\\
        }
       \KwOut{Optimal power allocation strategy $\boldsymbol{\rho}^{\ast} = \boldsymbol{\rho}^{[t]}$.}
\end{algorithm}

\vspace{-1.2em}

\subsection{Subproblem II: Rate-Splitting Optimization}

\vspace{-0.2em}

\par Given power allocation strategy $\overline{\boldsymbol{\rho}}\triangleq\left[\bar{\rho}_{c},\bar{\rho}_{p,1},\cdots,\bar{\rho}_{p,U}\right]^{T}$, $\widetilde{\mathcal{P}}2$ only depends on $\boldsymbol{\mathcal{R}}$ and $N_{T}$. Similarly, temporarily disregarding $N_{T}$, \textbf{Subproblem II} can be formulated as follows:

\vspace{-1.5em}

\begin{subequations}\label{e47}
   \begin{align}
      \widetilde{\mathcal{P}}2\mathrm{-II}:&\ \ \max_{\{\boldsymbol{\mathcal{R}}\}} \mathcal{T}\big(\boldsymbol{\mathcal{R}},\overline{\boldsymbol{\rho}}\big) = \mathcal{T}_{c}\big(\boldsymbol{\mathcal{R}}_{c}\big) + \mathcal{T}_{p}\big(\boldsymbol{\mathcal{R}}_{p}\big), \label{e47a}\\
      \quad s.t. & \quad (\ref{e23g}), (\ref{e23j}), (\ref{e38b})-(\ref{e38d}), (\ref{e39b})
   \vspace{-1em}
   \end{align}
\end{subequations}
where $\!\! \boldsymbol{\mathcal{R}}_{c} \!\! \triangleq \!\! \left[\mathcal{R}_{c},\mathcal{R}_{c,1},\cdots,\mathcal{R}_{c,U}\right]^{T}\!\!$, $\boldsymbol{\mathcal{R}}_{p} \!\! \triangleq \!\! \left[\mathcal{R}_{p,1},\cdots,\mathcal{R}_{c,U}\right]^{T}$, and

\vspace{-1.0em}

\begin{small}
\begin{subequations}\label{e48}
   \begin{align}
      & \mathcal{T}_{c}\big(\boldsymbol{\mathcal{R}}_{c}\big) = \sum_{u=1}^{U} \mathcal{R}_{c,u}\big(1-\varepsilon_{c,u}(\overline{\boldsymbol{\rho}},\mathcal{R}_{c})\big), \label{e48a}\\
      & \mathcal{T}_{p}\big(\boldsymbol{\mathcal{R}}_{p}\big) = \sum_{u=1}^{U} \mathcal{R}_{p,u}\big(1-\varepsilon_{p,u}(\overline{\boldsymbol{\rho}},\mathcal{R}_{p,u})\big). \label{e48b}
   \end{align}
\end{subequations}
\end{small}

\vspace{-0.4em}

\par Since $\varepsilon_{c,u}$ and $\varepsilon_{p,u}$ are only dependent to $\mathcal{R}_{c}$ and $\mathcal{R}_{p,u}$, (\ref{e38a}) and (\ref{e38b}) can be transformed into linear constraints. From $\varepsilon_{c,u} \leq \varepsilon_{th}$ and $\varepsilon_{p,u} \leq \varepsilon_{th}$, we can obtain that

\vspace{-1.2em}

\begin{subequations}\label{e49}
   \begin{align}
      & 0 \!<\! \mathcal{R}_{c} \! \leq \! \min_{\{u \in \mathcal{U}\}}\!\left\{C_{c,u}\big(N_{d},\overline{\boldsymbol{\rho}},\varepsilon_{th}\big)\right\} \!=\! \mathcal{R}_{c}^{\uparrow},\ \ \ \ \ \ \ \label{e49a}\\
      & 0 \!<\! \mathcal{R}_{p,u} \! \leq \! \sum\limits_{j=1}^{N_{R}}\!\!\bigg(\!\!\log_{2}\!\big(1\!+\!\Gamma_{p,u}^{(j)}\big)\!-\!\sqrt{\!\frac{\mathcal{V}(\Gamma_{p,u}^{(j)})}{N_{d}}}\frac{Q^{-1}(\varepsilon_{th})}{\ln2}\!\bigg)\nonumber\\
      & \overset{(a)}{=} N_{R}\bigg(\!\!\log_{2}\!\big(1\!+\!\overline{\Gamma}_{p,u}\big)\!-\!\sqrt{\!\frac{\mathcal{V}(\overline{\Gamma}_{p,u})}{N_{d}}}\frac{Q^{-1}(\varepsilon_{th})}{\ln2}\!\bigg) \! = \! \mathcal{R}_{p,u}^{\uparrow},\ \ \ \ \ \ \ \label{e49b}
\vspace{-0.5em}
   \end{align}
\end{subequations}
where $(a)$ can be derived from Theorem 1. %, and $\overline{\Gamma}_{p,u} = \frac{\bar{\rho}_{p,u}}{\sum_{k=1,k \neq u}\bar{\rho}_{p,k}}$.
Since $\widetilde{\mathcal{P}}2\mathrm{-II}$ remains a non-convex problem, directly solving it is challenging. Observing $\widetilde{\mathcal{P}}2\mathrm{-II}$, $\mathcal{T}_{c}\big(\boldsymbol{\mathcal{R}}_{c}\big)$ and $\mathcal{T}_{p}\big(\boldsymbol{\mathcal{R}}_{p}\big)$ are only coupled in (\ref{e23g}). In this case, we propose to decouple $\widetilde{\mathcal{P}}2\mathrm{-II}$ leveraging sequential optimization technique as follows:

\vspace{-1.2em}

\begin{subequations}\label{e50}
   \begin{align}
      \widetilde{\mathcal{P}}5:&\ \ \boldsymbol{\mathcal{R}}_{p}^{\ast} = \arg\max_{\{\boldsymbol{\mathcal{R}}_{p}\}} \mathcal{T}_{p}\big(\boldsymbol{\mathcal{R}}_{p}\big), \label{e50a}\\
      \quad s.t. & \quad (\ref{e49b}). \label{e50b}
   \vspace{-1em}
   \end{align}
\end{subequations}

\vspace{-1.5em}

\begin{subequations}\label{e51}
   \begin{align}
      \widetilde{\mathcal{P}}6:&\ \ \boldsymbol{\mathcal{R}}_{c}^{\ast} = \arg\max_{\{\boldsymbol{\mathcal{R}}_{c}\}} \mathcal{T}_{c}\big(\boldsymbol{\mathcal{R}}_{c}\big), \label{e51a}\\
      \quad s.t. & \quad \mathcal{R}_{c,u} \geq \mathcal{R}_{min} - \mathcal{R}_{p,u}^{\ast}, \label{e51b}\\
                 & \quad (\ref{e49a}). \label{e51c}
   \vspace{-1em}
   \end{align}
\end{subequations}

\vspace{-0.5em}

\par According to sequential optimization techniques, $\widetilde{\mathcal{P}}5$ must be tackled first, and then proceed to solve $\widetilde{\mathcal{P}}6$ using the results obtained from $\widetilde{\mathcal{P}}5$. \textbf{Lemma 4} shows that $\widetilde{\mathcal{P}}5$ is a strictly convex problem that is solved through interior-point methods.

\vspace{-0.4em}

\begin{lemma}\label{le4}
   $\widetilde{\mathcal{P}}5$ is a strictly convex problem in $\left(0,\mathcal{R}_{p,u}^{\uparrow}\right)$.
\end{lemma}

\vspace{-1.2em}

\begin{proof}
   The proof of \textbf{Lemma 4} is given in Appendix E.
\end{proof}

\vspace{-0.7em}

\par Subsequently, we focus on addressing $\widetilde{\mathcal{P}}6$. Combining (\ref{e48a}), and (\ref{e51a})-(\ref{e51c}), it can be readily observed that if $\mathcal{R}_{c}$ is given, the objective function (\ref{e48a}) of $\widetilde{\mathcal{P}}6$ becomes a linear function of $\mathcal{R}_{c,u}$, in this case, we can obtain

\vspace{-1.5em}

\begin{subequations}\label{e52}
   \begin{align}
      \widetilde{\mathcal{P}}7:& \quad \max_{\{\mathcal{R}_{c,u}\}_{u \in \mathcal{U}}} \sum\limits_{u \in \mathcal{U}}\widetilde{\mathcal{W}}_{c,u}\mathcal{R}_{c,u} , \nonumber\\
      \quad s.t. & \quad \mathcal{R}_{c,u} \geq \mathcal{R}_{min} - \mathcal{R}_{p,u}^{\ast}, \nonumber\\
                 & \quad \sum\limits_{u \in \mathcal{U}}\mathcal{R}_{c,u} = \mathcal{R}_{c}, \nonumber
   \vspace{-1.5em}
   \end{align}
\end{subequations}
where $\widetilde{\mathcal{W}}_{c,u} = 1 - \varepsilon_{c,u}\left(\overline{\boldsymbol{\rho}},\mathcal{R}_{c}\right)$. $\widetilde{\mathcal{P}}7$ is a standard linear programming problem, whose optimal solutions are derived leveraging the Lagrange multiplier method, denoted by

\vspace{-0.5em}

\begin{equation}\label{e53}
   \mathcal{R}_{c,u}^{\ast} = \left\{
                                \begin{array}{ll}
                                  \mathcal{R}_{c,u}^{max}, & \hbox{if $\overline{\Gamma}_{c,u} = \max\limits_{k \in \mathcal{U}}\{\overline{\Gamma}_{c,k}\}$;} \\
                                  \mathcal{R}_{c,u}^{min}, & \hbox{otherwise,}
                                \end{array}
                              \right.
\end{equation}
where $\mathcal{R}_{c,u}^{min} \!=\! \max\{\mathcal{R}_{min} \!-\! \mathcal{R}_{p,u}^{\ast},0\}$ and $\mathcal{R}_{c,u}^{max} \!=\! \max\{\mathcal{R}_{c} \!-\! \mathcal{R}_{p,u}^{\ast},0\}$. Then, $\widetilde{\mathcal{P}}6$ can be reformulated by using (\ref{e53}), as follow:

\vspace{-1.2em}

\begin{small}
\begin{subequations}\label{e54}
   \begin{align}
      \widetilde{\mathcal{P}}8:&\ \ \mathcal{R}_{c}^{\ast} = \arg\max_{\{\mathcal{R}_{c}\}}\bigg\{\sum\limits_{k=1,k\neq u}\mathcal{R}_{c,k}^{min}\left(1-\varepsilon_{c,k}(\mathcal{R}_{c})\right) + \nonumber \\
                 & \bigg(\mathcal{R}_{c} - \sum\limits_{k=1,k\neq u}\mathcal{R}_{c,k}^{min}\bigg)\big(1-\varepsilon_{c,k}(\mathcal{R}_{c})\big)\bigg\}, \label{e54a}\\
      \quad s.t. & \quad 0 < \mathcal{R}_{c} \leq \mathcal{R}_{c}^{\uparrow}. \label{e54b}
   \vspace{-1em}
   \end{align}
\end{subequations}
\end{small}

\vspace{-0.2em}

\par \textbf{Lemma 5} provides evidence that $\widetilde{\mathcal{P}}8$ is a strictly convex problem, which can be effectively addressed through standard convex optimization algorithms.

\vspace{-0.5em}

\begin{lemma}\label{le5}
   $\widetilde{\mathcal{P}}8$ is a strictly convex problem in $\left(0,\mathcal{R}_{c}^{\uparrow}\right)$.
\end{lemma}

\vspace{-1.4em}

\begin{proof}
   The proof of \textbf{Lemma 5} is given in Appendix F.
\end{proof}

\vspace{-1.2em}

\subsection{Joint Power Allocation, Rate Splitting, and Transmit Antenna Selection Optimization Mechanism}

\vspace{-0.3em}
\par Increasing transmit antennas leads to improvements in system performance, capacity, and reliability. Meanwhile, it results in escalated hardware costs, power consumption, complex signal processing, and heightened sensitivity to interference and DEP. As a result, transmit antenna selection becomes a trade-off problem involving convex integer optimization, which can be efficiently solved by the one-dimensional integer search method \cite{10382447}. Combining Sec. III-D and E, we propose the joint power allocation, rate splitting, and transmit antenna selection (JPRT) optimization algorithm, the detailed description of which is provided in \textbf{Algorithm 2}.

\vspace{-0.6em}

\subsection{Optimality and Complexity Analysis}

\vspace{-0.2em}

\par To analyze the optimality of the proposed JPRT optimization algorithm, \textbf{Theorem 2} is derived as follows:

\vspace{-0.4em}

\begin{theorem}\label{theo2}
   The proposed JPRT optimization algorithm achieves a progressively improved total ETR in each iteration and eventually converges to a local optimum.
\end{theorem}

\vspace{-0.9em}

\begin{proof}
    As illustrated in Fig. 2, we first analyze the convergence of steps $\textbf{4} \!\!\sim\!\! \textbf{15}$ in \textbf{Algorithm 2}. In the $t$-th iteration, we have $N_{T}^{\ast}[n]$. In the case of a given $\!\boldsymbol{\rho}^{[t]}[n]$, $\mathcal{T}\!\left(\!N_{T}^{\ast}[n],\boldsymbol{\mathcal{R}}^{[t]}[n],\boldsymbol{\rho}^{[t]}[n]\!\right)\![n]$ only depends on $\boldsymbol{\mathcal{R}}^{[t]}[n]$. After performing rate-splitting optimization, we can obtain

\vspace{-1.6em}

\begin{equation}\label{e55}
    \mathcal{T}\!\!\left(\!N_{T}^{\ast}[n],\!\boldsymbol{\mathcal{R}}^{\ast[t]}[n],\boldsymbol{\rho}^{[t]}[n]\!\right)\![n] \!\geq\! \mathcal{T}\!\!\left(\!N_{T}^{\ast}[n],\boldsymbol{\mathcal{R}}^{[t]}[n],\boldsymbol{\rho}^{[t]}[n]\!\right)\![n],
\vspace{-0.3em}
\end{equation}
where $\boldsymbol{\mathcal{R}}^{\ast[t]}[n]$ indicates the optimal rate-splitting strategy for a given $\boldsymbol{\rho}^{[t]}[n]$ in the $t$-th iteration. On the other hand, we can also obtain that

\vspace{-1.3em}

\begin{small}
\begin{equation}\label{e56}
  \!\!\!\!\!\!\!\! \mathcal{T}^{\ast}\!\!\left(\!N_{T}^{\ast}[n],\boldsymbol{\mathcal{R}}^{\ast[t]}[n],\boldsymbol{\rho}^{\ast[t]}[n]\!\right)\![n] \!\geq\! \mathcal{T}\!\!\left(\!N_{T}^{\ast}[n],\boldsymbol{\mathcal{R}}^{\ast[t]}[n],\boldsymbol{\rho}^{[t]}[n]\!\right)\![n].
\end{equation}
\end{small}

\vspace{-0.5em}

\par Based on the results obtained in the $t$-th iteration, we can derive that

\vspace{-1.3em}

\begin{small}
\begin{equation}\label{e57}
  \mathcal{T}\!\!\left(\!N_{T}^{\ast}[n],\boldsymbol{\mathcal{R}}^{\ast[t+1]}[n],\boldsymbol{\rho}^{\ast[t]}[n]\!\right)\![n] \!\geq\! \mathcal{T}^{\ast}\!\!\left(\!N_{T}^{\ast}[n],\boldsymbol{\mathcal{R}}^{\ast[t]}[n],\boldsymbol{\rho}^{\ast[t]}[n]\!\right)\![n].
\end{equation}
\end{small}

\par After performing power allocation optimization, we can obtain that

\vspace{-1.2em}

\begin{small}
\begin{equation}\label{e58}
   \begin{aligned}
     & \mathcal{T}^{\ast}\!\! \left(\!N_{T}^{\ast}[n],\boldsymbol{\mathcal{R}}^{\ast[t+1]}[n],\boldsymbol{\rho}^{\ast[t+1]}[n]\!\right)\![n]\\
     & \quad \quad \quad \quad \quad \quad \quad \quad \quad \quad \quad \geq\! \mathcal{T}\!\!\left(\!N_{T}^{\ast}[n],\boldsymbol{\mathcal{R}}^{\ast[t+1]}[n],\boldsymbol{\rho}^{\ast[t]}[n]\!\right)\![n].
   \end{aligned}
\end{equation}
\end{small}

\vspace{-0.4em}

\par After repeated iterations, the total ETR progressively increases in each alternating iteration. Ultimately, the total ETR will converge towards a local optimum and reach $\mathcal{T}^{\ast}\!\left(N_{max}\right)\!-\!\mathcal{T}^{\ast}\!\left(N_{max}-1\right) \!\!\leq\!\! \Delta_{th}$, where $N_{max}$ and $\Delta_{th}$ denote the number of iterations and the convergence criteria, respectively.

\end{proof}

\vspace{-1em}

\begin{figure}[htbp]
\vspace{-1em}
\centering
\includegraphics[scale=0.66]{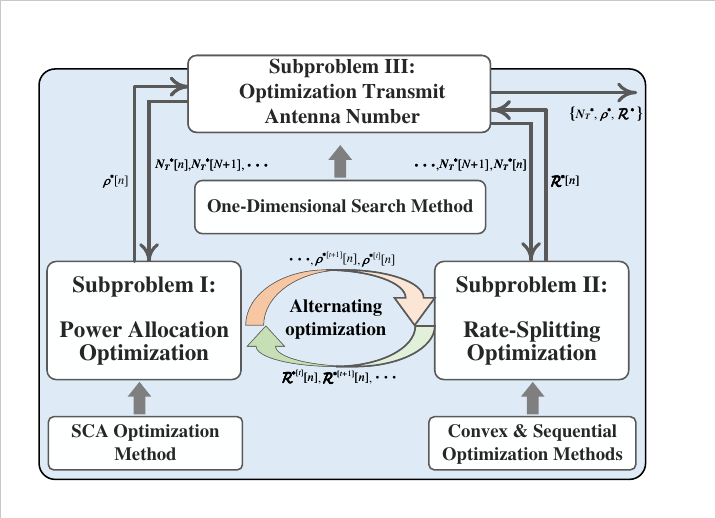}
\vspace{-0.2em}
\caption{The illustration of the proposed JPRT optimization mechanism.}
\label{fig:label}
\vspace{-1.0em}
\end{figure}

\par The computational complexity of the proposed JPRT optimization algorithm comprises three parts. The first part is power allocation, solved by leveraging the SCA technique, with complexity $\mathcal{O}\!\left(N_{1}Q\right)$, where $Q \!=\! \left(7+N_{T}\right)U$ indicates the number of optimization variables updated during each iteration, and $N_{1} \!=\! \sqrt{2U+1}\log_{2}\!\left(1/\epsilon\right)$ signifies the required iterations \cite{marks1978general,li2012coordinated}. The second part is rate-splitting optimization, where $\widetilde{\mathcal{P}}_{5}$ and $\widetilde{\mathcal{P}}_{6}$ are tackled through interior-point methods, with complexity $\mathcal{O}\left(U^{3.5}\log_{2}\!(1/\epsilon)\right)$ \cite{wright1997primal}. Finally, the transmit antenna selection optimization, executed post-alternating optimization iterations, is tackled through one-dimensional search methods with complexity $\mathcal{O}\left(log_{2}\left(N_{T}^{max}\right)\right)$ \cite{tsai2010golden, 10382447}. Therefore, the total computational complexity of \textbf{Algorithm 2} can be denoted as $\mathcal{O}\!\bigg(\!\sum\limits_{i=1}^{N_{max}}\!N_{i}\sqrt{2}N_{T}^{*3}U^{3.5}\log_{2}\!(1/\epsilon)\!\!\bigg)$, where $N_{i}$ and $N_{max}\sim \mathcal{O}\left(log_{2}\left(N_{T}^{max}\right)\right)$ are iteration numbers for alternating optimization and transmit antenna selection, respectively.

\begin{algorithm}[h]% [htbp]
\small
 \setstretch{0.75}
        \caption{JPRT Optimization Algorithm.}
        \KwIn{Lower search bound $N_{T}^{min} = 0$; Upper search bound $N_{T}^{max} = N_{tot}$;}
        \textbf{Initialize:} Iteration index $n = 1$; $LB[1] = N_{T}^{min}$; $UB[1] = N_{T}^{max}$;\\
        \While{\text{\rm{Convergence-1 Unsatisfied}}}{
           Set $N_{T}^{\ast}[n] = \lfloor\left(LB[n] + UB[n]\right)/2\rfloor$ ;\\
           \textbf{Initialize:} $\overline{\boldsymbol{\mathcal{R}}}^{\ast[0]}[n]$, iteration index $t = 0$;\\
           \While{\text{\rm{Convergence-2 Unsatisfied}}}{
              \textcolor[rgb]{1.00,0.00,0.50}{\tcp{\textbf{$\boldsymbol{\widetilde{\mathcal{P}}2}$-I: Power Allocation Optimization}}}
              Input $\overline{\boldsymbol{\mathcal{R}}}^{\ast[t]}[n]$ and $N_{T}^{\ast}[n]$ into \textbf{Algorithm 1};\\
              Execute Algorithm 1 to obtain $\boldsymbol{\rho}^{\ast[t]}[n]$;\\
              \textcolor[rgb]{1.00,0.00,0.50}{\tcp{\textbf{$\boldsymbol{\widetilde{\mathcal{P}}2}$-II: Rate-Splitting Optimization}}}
              Input $\boldsymbol{\rho}^{\ast[t]}[n]$ and $N_{T}^{\ast}[n]$ into $\boldsymbol{\widetilde{\mathcal{P}}2}$-II;\\
              Solve convex problem $\widetilde{\mathcal{P}}5$ to obtain $\mathcal{R}_{p,u}^{\ast[t+1]}[n], \forall u \in \mathcal{U}$;\\
              Solve convex problem $\widetilde{\mathcal{P}}8$ to obtain $\mathcal{R}_{c}^{\ast[t+1]}[n]$;\\
              Substitute $\mathcal{R}_{c}^{\ast[t+1]}[n]$ into $\widetilde{\mathcal{P}}7$ and solve to obtain $\mathcal{R}_{c,u}^{\ast[t+1]}[n], \forall u \in \mathcal{U}$;\\
              Set $\boldsymbol{\mathcal{R}}^{\ast[t+1]}[n] = \left[\mathcal{R}_{c}^{\ast[t+1]}[n],\!\mathcal{R}_{c,u}^{\ast[t+1]}[n],\mathcal{R}_{p,u}^{\ast[t+1]}[n]\right]^{T}\!\!, \forall u \in \mathcal{U}$;\\
              Set $t = t + 1$;\\
             }
              Set $\boldsymbol{\rho}^{\ast}[n]\!=\!\boldsymbol{\rho}^{\ast[t]}[n]$ and $\boldsymbol{\mathcal{R}}^{\ast}[n] \!=\! \boldsymbol{\mathcal{R}}^{\ast[t+1]}[n]$;\\
             \textcolor[rgb]{1.00,0.00,0.50}{\tcp{\textbf{Transmit Antenna Selection Optimization}}}
             \eIf{\text{\rm{Convergence-2 Unsatisfied}}}{
                Set $LB[n+1] = \lfloor\left(LB[n] + UB[n]\right)/2\rfloor$ ;\\
             }
             {
                Set $UB[n+1] = \lfloor\left(LB[n] + UB[n]\right)/2\rfloor$ ;\\
             }
                Set $n = n+1$;\\
        }
       \KwOut{Optimal Solution $\boldsymbol{\rho}^{\ast}[n], \boldsymbol{\mathcal{R}}^{\ast}[n], N_{T}^{\ast}[n]$.}
\end{algorithm}

\vspace{-2em}

\section{Performance Evaluation}
\par In this section, extensive numerical results are presented to investigate and demonstrate the superiority, distinctive features, and expansive applicability of the developed RSMA-mMIMO-xURLLC network architecture. In particular, meticulous scrutiny is conducted to assess the technological advancement exhibited by the proposed JPRT optimization algorithm, and several state-of-the-art multiple access techniques are exploited to perform a comprehensive performance comparison. We consider $U$ UEs uniformly distributed within a ring-shaped area around the BS, with distances ranging from $R_{min} = 35$ \emph{m} to $R_{max} = 95$ \emph{m}. The small-scale fading is modeled as Rayleigh fading, while the large-scale fading coefficient $\kappa_{u}$ can be modeled as $\kappa_{u} = [\lambda/(4\pi d_{u})]^{2}$, where $d_{u}$ represents the distance of UE $u \in \mathcal{U}$ from the BS, uniformly distributed within $(R_{min}, R_{max})$. Unless explicitly stated otherwise, other simulation parameters are presented in Table I.
\begin{table}[t]
 \renewcommand{\arraystretch}{1.0}
 \caption{Simulation Parameter Settings}
 \vspace{-0.5em}
 \label{table_II}
 \centering
 \resizebox{0.73\columnwidth}{!}
 {
 \begin{tabular}{|c|l|c|c|l|c|p{0.3cm}}
  \hline
  \bfseries Parameter & \bfseries Physical meaning &\bfseries Value\\
  \hline
  $B_{tot}$ & System bandwidth & $1$ MHz\\
  \hline
  $\lambda = \frac{c}{f_{c}}$ & Wavelength & $15$ cm \\
  \hline
  $f_{c}$ & Carrier frequency & $2$ GHz\\
  \hline
  $\epsilon$ & Approximate error & $10^{-10}$\\
% M. Grant, S. Boyd, and Y. Ye. (2008). CVX: MATLAB Software for Disciplined Convex Programming. [Online]. Available: http://cvxr. com/cvx/citing/
  \hline
  $\sigma^{2}$ & Power of noise & $-113$ dBm\\
  \hline
  $D_{th}$ & Latency constraint & $1$ ms\\
  \hline
  $\varepsilon_{th}$ & Reliability (Error decoding) constraint & $10^{-5}$\\
  \hline
  $R_{min}$ & Minimum distance & $35$ m\\
  \hline
  $R_{max}$ & Maximum distance & $95$ m\\
  \hline
 \end{tabular}
 }
\end{table}

\begin{figure*}[h]
  \centering
  \begin{minipage}[b]{0.3\textwidth}
    \includegraphics[scale=0.25]{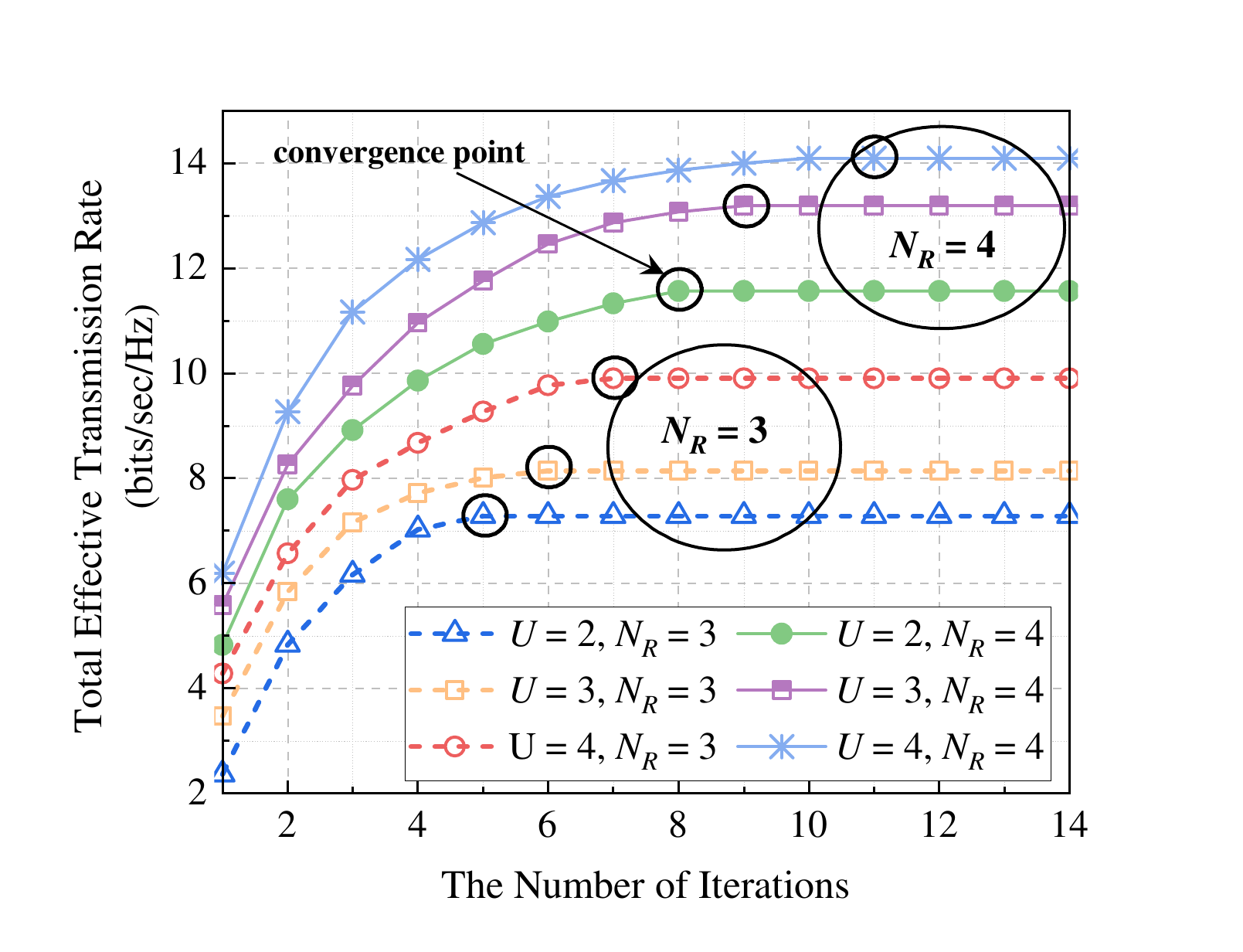}
    \caption{The convergence behavior of Algorithm 2. $P_{tot} = 4$ W, $N_{tot} = 10^{3}$ CUs, and $\mathcal{R}_{min} = 2.0$ bits/sec/Hz.}
    \label{fig3}
  \end{minipage}
  \hfill
  \begin{minipage}[b]{0.3\textwidth}
    \includegraphics[scale=0.25]{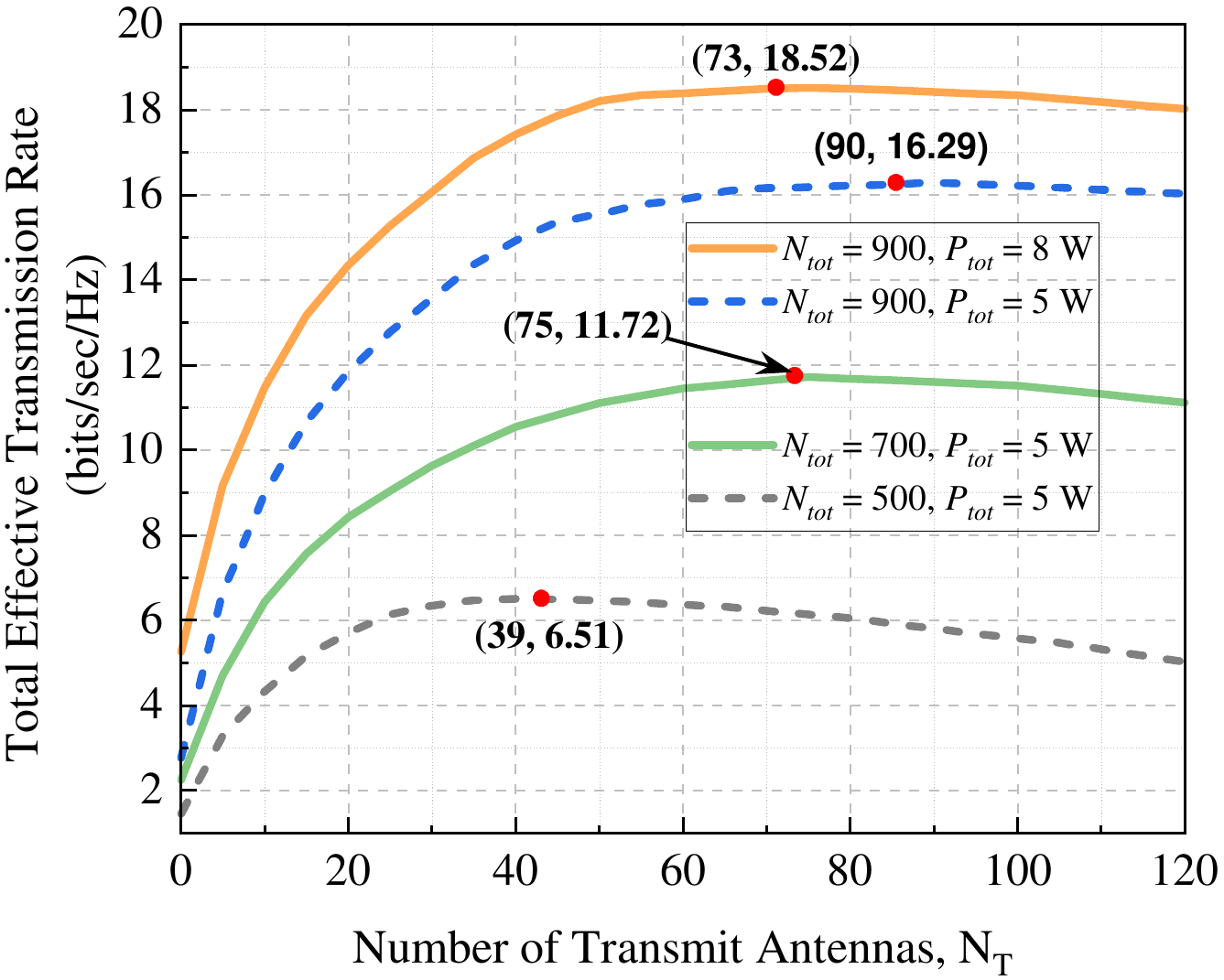}
    \caption{The optimality performance of Algorithm 2. $U = 5$, $N_{R} = 4$, and $\mathcal{R}_{min} = 2.0$ bits/sec/Hz.}
    \label{fig4}
  \end{minipage}
  \hfill
  \begin{minipage}[b]{0.3\textwidth}
    \includegraphics[scale=0.25]{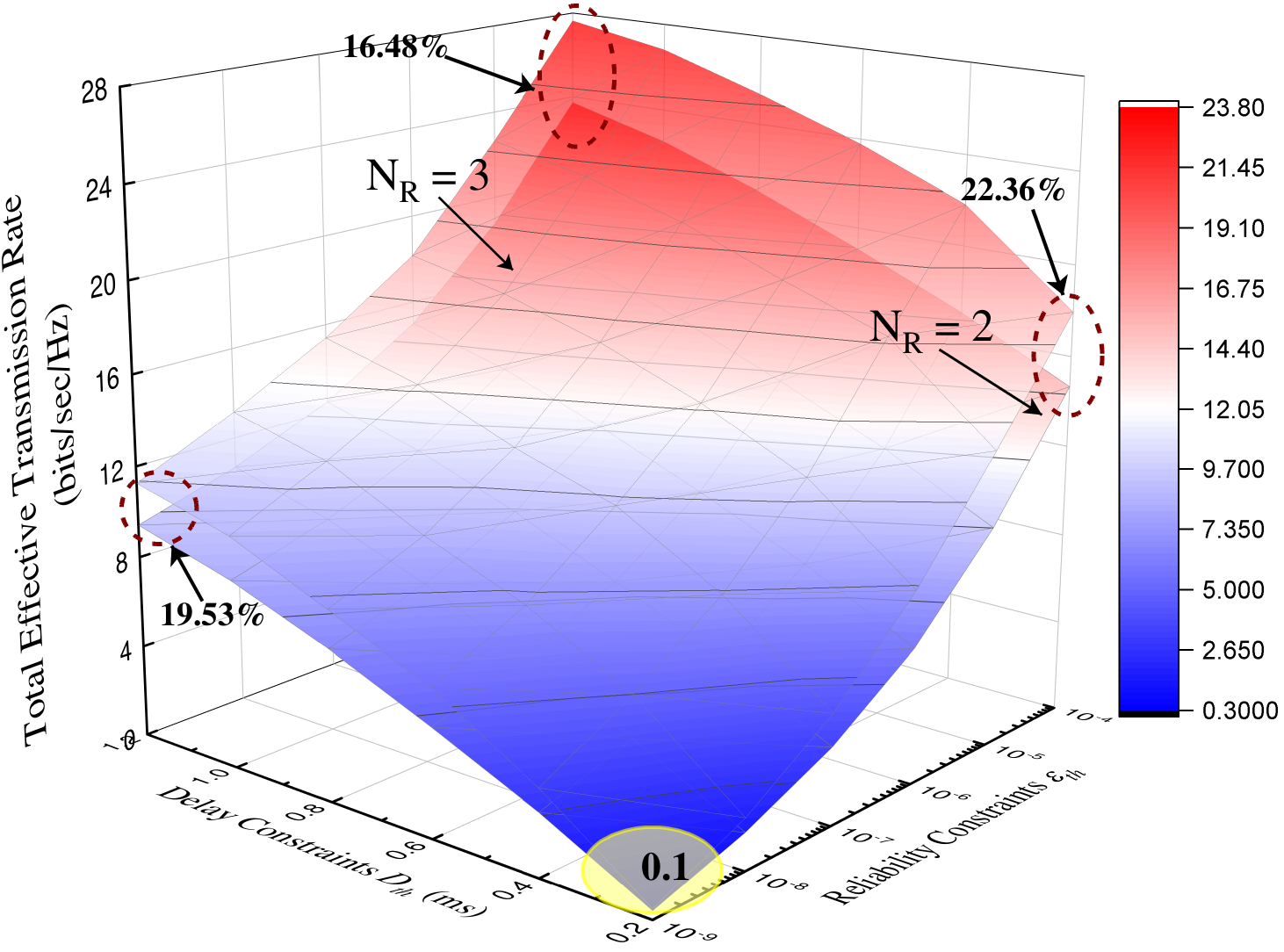}
    \caption{Maximum ETR versus delay and reliability constraints. $U = 5$, $P_{tot} = 5$ W, and $\mathcal{R}_{min} = 2.0$ bits/sec/Hz.}
   \label{fig5}
  \end{minipage}
  \vspace{-1.5em}
\end{figure*}

\subsection{Convergence and Optimality Analysis}

\vspace{-0.5em}

\par Fig. \ref{fig3} illustrates the convergence behavior of the proposed JPRT optimization algorithm. With each iteration, the maximum total ETR progressively increases and rapidly converges, demonstrating \textbf{Algorithm 2}'s superior convergence performance and aligning with \textbf{Theorem 2}'s theoretical analysis. By maintaining $N_{R}$ while increasing the number of UEs, the maximum total ETR can be improved. As $N_{R}$ increases, the maximum total ETR also experiences remarkable gains. These improvements primarily stem from two perspectives. Firstly, our network architecture enables the messages designated for each receiver to be split into common and private parts, and mMIMO systems inherently possess exceptional interference suppression capabilities. Secondly, increasing the number of receiving antennas within mMIMO systems can effectively enhance spatial diversity, which effectively enhances the SINRs of the received signals, thereby improving the total ETR.

\par As depicted in Fig. \ref{fig4}, the impact of the transmit antenna selection $N_{T}$ on the maximum total ETR is examined. It can be observed that a discernible tradeoff exists between the total ETR and $N_{T}$. Firstly, as $N_{T}$ increases, the maximum total ETR increases to a peak before gradually declining. The apex of this relationship is marked with pink asterisks $\ast$ in Fig. \ref{fig4}. When the total transmit power $P_{tot} = 5 W$, the optimal $N_{T}^{*}$ increases with blocklength $N_{tot}$. This is primarily attributed to the constrained power and bandwidth allocated to each CU, and an increase in $N_{tot}$ leads to more intensive utilization of wireless resources, thereby necessitating more $N_{T}$ to sustain the required SINR and fulfill xURLLC's QoS requirements. Therefore, increasing $N_{T}$ is beneficial to guarantee stability and reliability during short-packet transmission. Conversely, at $N_{tot} = 900$, the optimal $N_{T}^{*}$ decreases with increasing $P_{tot}$. This is mainly because with the increase of $P_{tot}$, SINR can be improved, allowing the system to liberate more wireless resources and reduce the number of CUs dedicated to pilot training, thus allocating more channel uses to be allocated to data transmission phase. Consequently, the reduction of the optimal $N_{T}^{*}$ can effectively utilize available channel resources, thereby enhancing the total ETR.

\par Fig. \ref{fig5} illustrates the relationship between maximum total ETR and xURLLC's QoS requirements $\left(D_{th},\varepsilon_{th}\right)$. With fixed $\varepsilon_{th}$, the maximum total ETR exhibits an upward trend with the enlargement of $D_{th}$, albeit at a decreasing rate. Similarly, for constant $D_{th}$, the maximum total ETR declines rapidly as $\varepsilon_{th}$ decreases, though the descent gradually tapers off. This primarily stems from the fact that as $\left(D_{th},\varepsilon_{th}\right)$ relaxs, more resources are available for short-packet data transmission. However, indiscriminately relaxing a particular QoS indicator cannot indefinitely enhance system performance due to strict constraints on blocklength and transmit power. This rationale can be underscored in Fig. \ref{fig5}, where a stringent reliability requirement (e.g., $\varepsilon_{th} \!=\! 10^{-9}$) limits performance gains despite latency requirement relaxation (e.g., $D_{th} \!=\! 0.2 \! \sim \! 1.2$ \emph{ms}).

\vspace{-0.4em}

\subsection{Comprehensive Performance Comparison}

\vspace{-0.3em}
\par To demonstrate the superior performance of the developed RSMA-mMIMO-xURLLC network architecture, the current state-of-the-art multiple access technologies, i.e., NOMA and SDMA, are selected as benchmarks. RSMA unifies NOMA and SDMA in terms of technical principles under downlink mMIMO scenarios as follows:

\begin{itemize}
  \item For the NOMA-assisted xURLLC, NOMA can be derived as a sub-scheme of RSMA, leveraging each common stream to fully encode the entire message for individual users \cite{mao2022rate}.
  \item For SDMA-assisted xURLLC, by prohibiting the power allocation process for all common streams, RSMA seamlessly transforms into SDMA, thereby making SDMA a sub-scheme of all linear precoding-based RSMA \cite{clerckx2024multiple}.
\end{itemize}

\begin{figure*}[t]
  \centering
  \begin{minipage}[b]{0.3\textwidth}
    \includegraphics[scale=0.25]{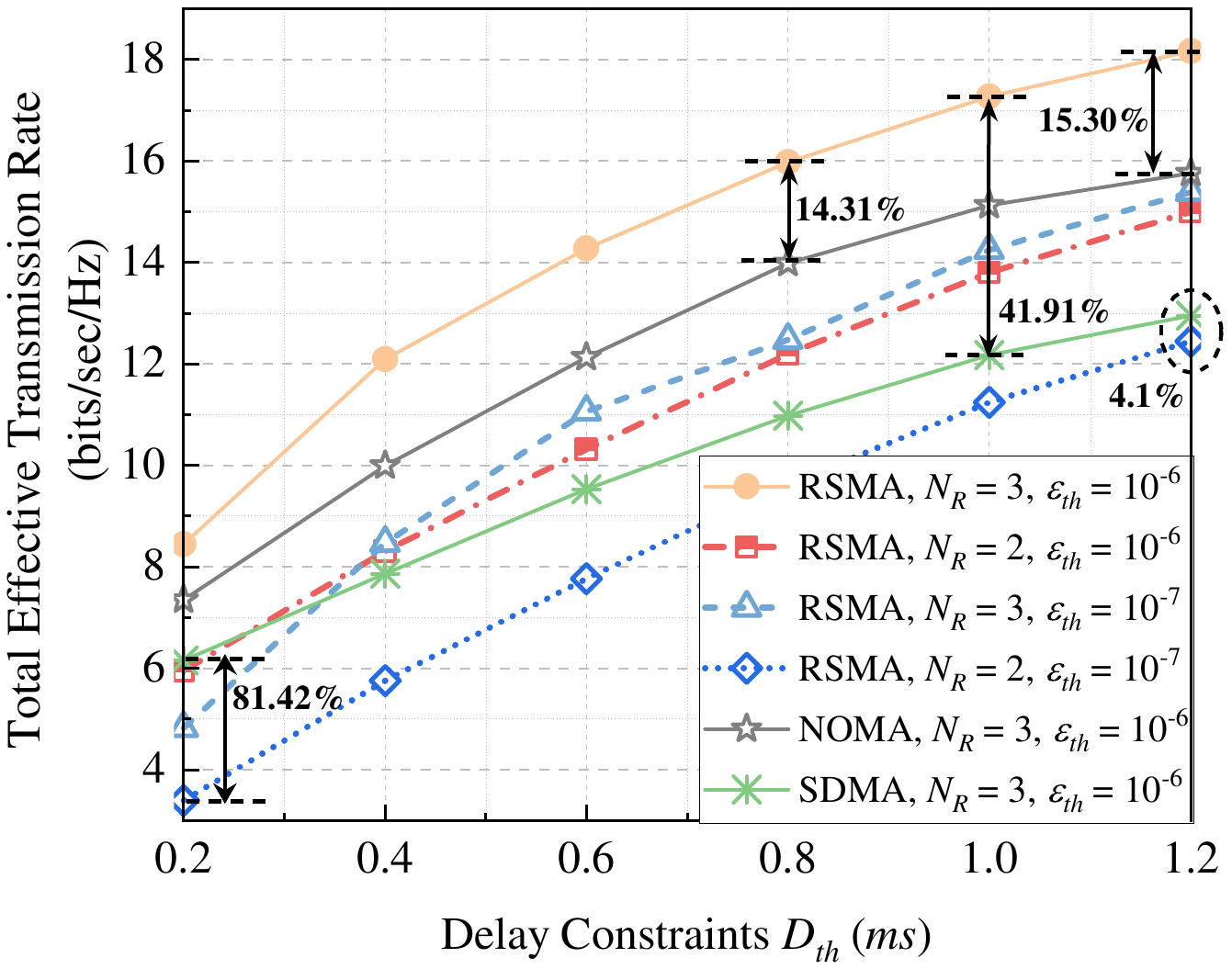}
    \caption{Maximum ETR versus delay constraints. $U = 5$, $P_{tot} = 5$ W, and $\mathcal{R}_{min} = 2.0$ bits/sec/Hz.}
    \label{fig6}
  \end{minipage}
  \hfill
  \begin{minipage}[b]{0.3\textwidth}
    \includegraphics[scale=0.25]{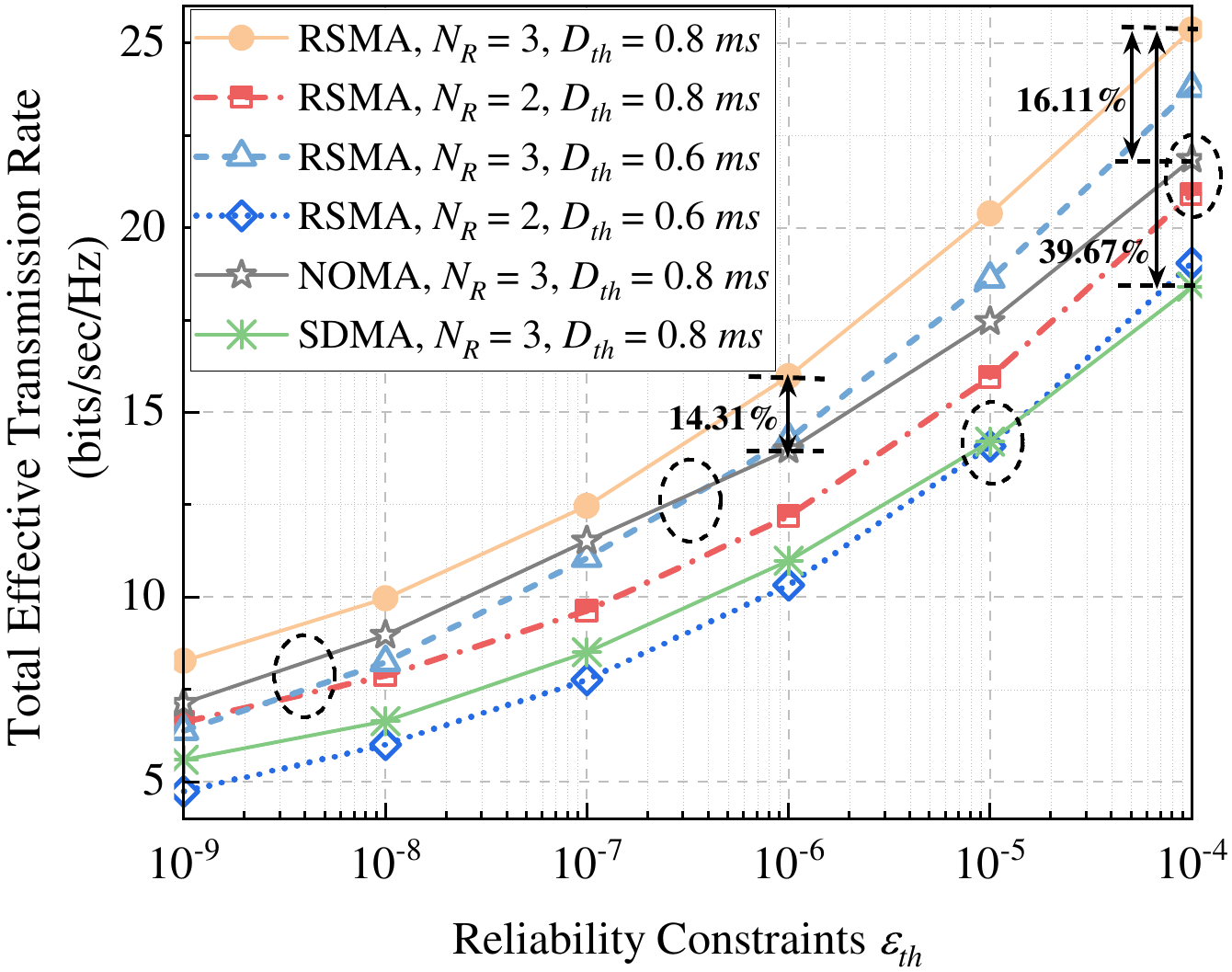}
    \caption{Maximum ETR versus reliability constraints. $U = 5$, $P_{tot} = 5$ W, and $\mathcal{R}_{min} = 2.0$ bits/sec/Hz.}
    \label{fig7}
  \end{minipage}
  \hfill
  \begin{minipage}[b]{0.3\textwidth}
    \includegraphics[scale=0.25]{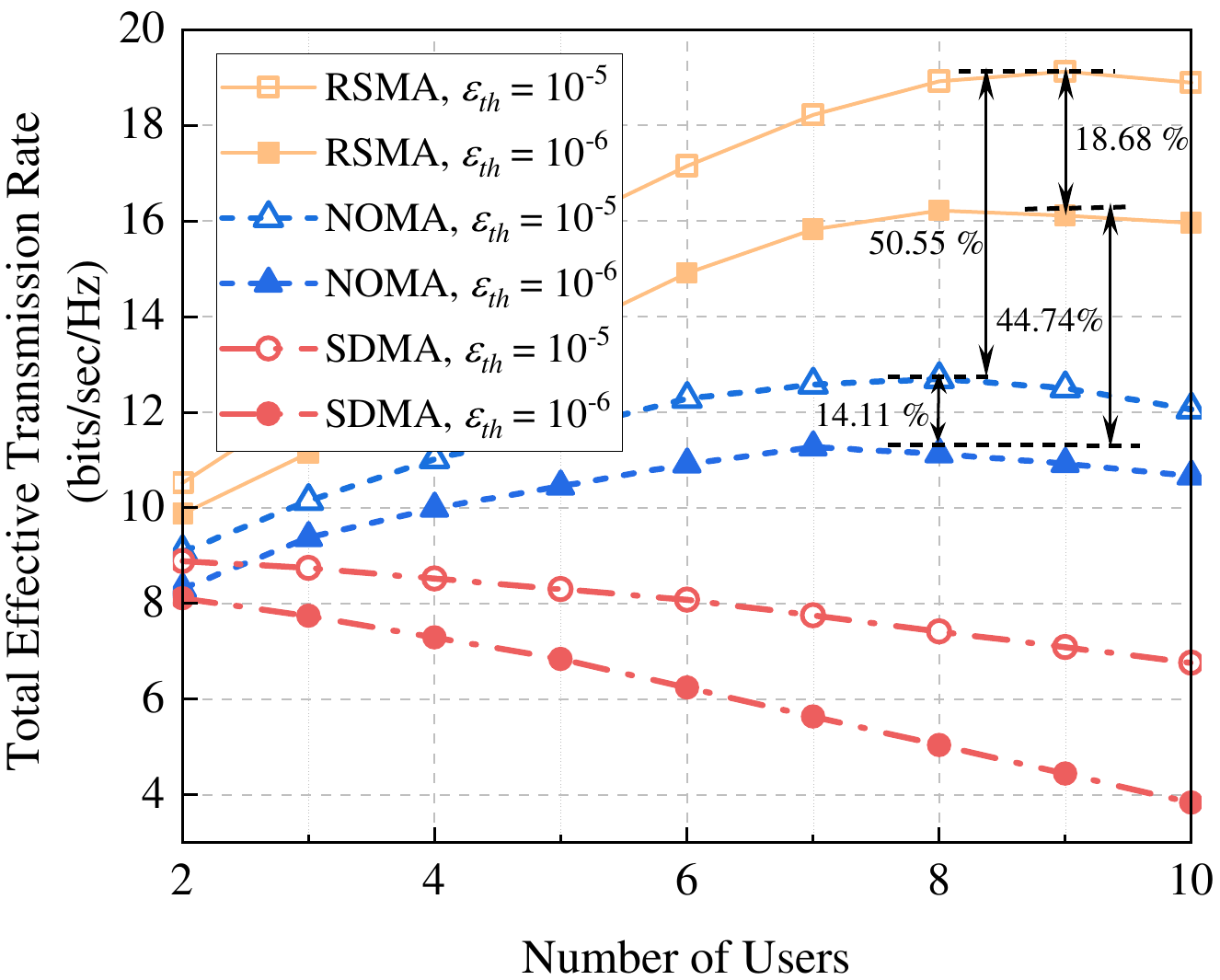}
    \caption{Relationship between the maximum total ETR and the number of accessed users. $P_{tot} = 5$ W, $N_{R} = 3$, and $\mathcal{R}_{min} = 2.0$ bits/sec/Hz.}
    \label{fig8}
  \end{minipage}
  \vspace{-1.5em}
\end{figure*}

\vspace{-0.4em}

\par As depicted in Fig. \ref{fig6}, the relationship between the maximum total ETR and the latency constraints is explored. At $N_{R} = 3$ and $\varepsilon_{th} = 10^{-6}$, our architecture outperforms NOMA and SDMA by approximately $15.3\%$ and $41.91\%$ performance enhancements, respectively, with enhancements growing as $D_{th}$ increases. This underscores our architecture's potential for increased throughput and reduced latency for xURLLC. As $D_{th}$ increases, our network architecture achieves performance gains comparable to NOMA, even under more stringent reliability requirements. This primarily stems from the critical discrepancy between RSMA and NOMA. RSMA can flexibly manage interference by partially decoding it and treating it as noise, softly bridging the two extremes of completely decoding interference and treating interference as noise. Notably, our architecture narrows the performance gaps with SDMA from $81.42\%$ to $4.2\%$ as Dth increases from $0.2$ \emph{ms} to $1.2$ \emph{ms} with fewer antennas and stricter reliability requirements. This is mainly because RSMA is the extension of NOMA and SDMA, and it seamlessly integrates with mMIMO systems, striking a balance between optimizing multiplexing and diversity gains. These numerical results substantiate the effectiveness of our architecture in providing enhanced QoS guarantees for xURLLC.

\par In Fig. \ref{fig7}, we explore the relationship between the maximum total ETR and reliability constraints. At $N_{R} = 3$ and $D_{th} = 0.8$ \emph{ms}, our architecture demonstrates around $16.11\%$ and $39.67\%$ performance improvements over NOMA and SDMA, respectively, underscoring its potential to guarantee more stringent reliability constraints for xURLLC. Moreover, as $\varepsilon_{th}$ increases, our architecture consistently outperforms NOMA, particularly in achieving lower latency, further highlighting its superiority over SDMA. Therefore, similar to the findings in Fig. \ref{fig6}, RSMA exhibits remarkable performance compared to NOMA and SDMA, positioning it as a pivotal technology in 6G networks, poised to meet the stringent QoS requirements.

\vspace{-0.4em}

\begin{figure*}[t]
  \centering
  \begin{minipage}[b]{0.3\textwidth}
    \includegraphics[scale=0.25]{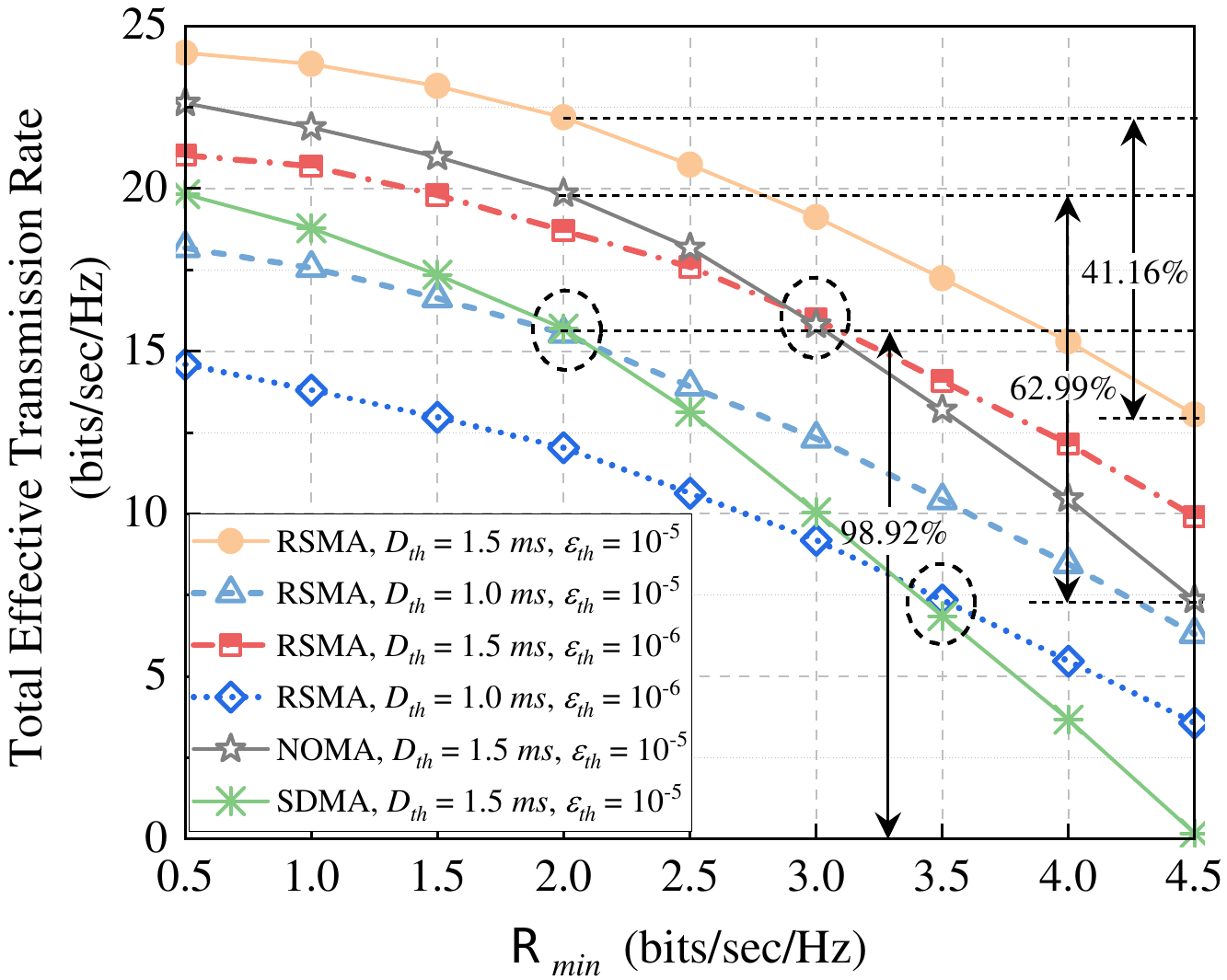}
    \caption{Relationship between the maximum total ETR and $\mathcal{R}_{min}$. $U = 5$, $N_{R} = 3$, and $P_{tot} = 5$ W}
    \label{fig9}
  \end{minipage}
  \hfill
  \begin{minipage}[b]{0.3\textwidth}
    \includegraphics[scale=0.25]{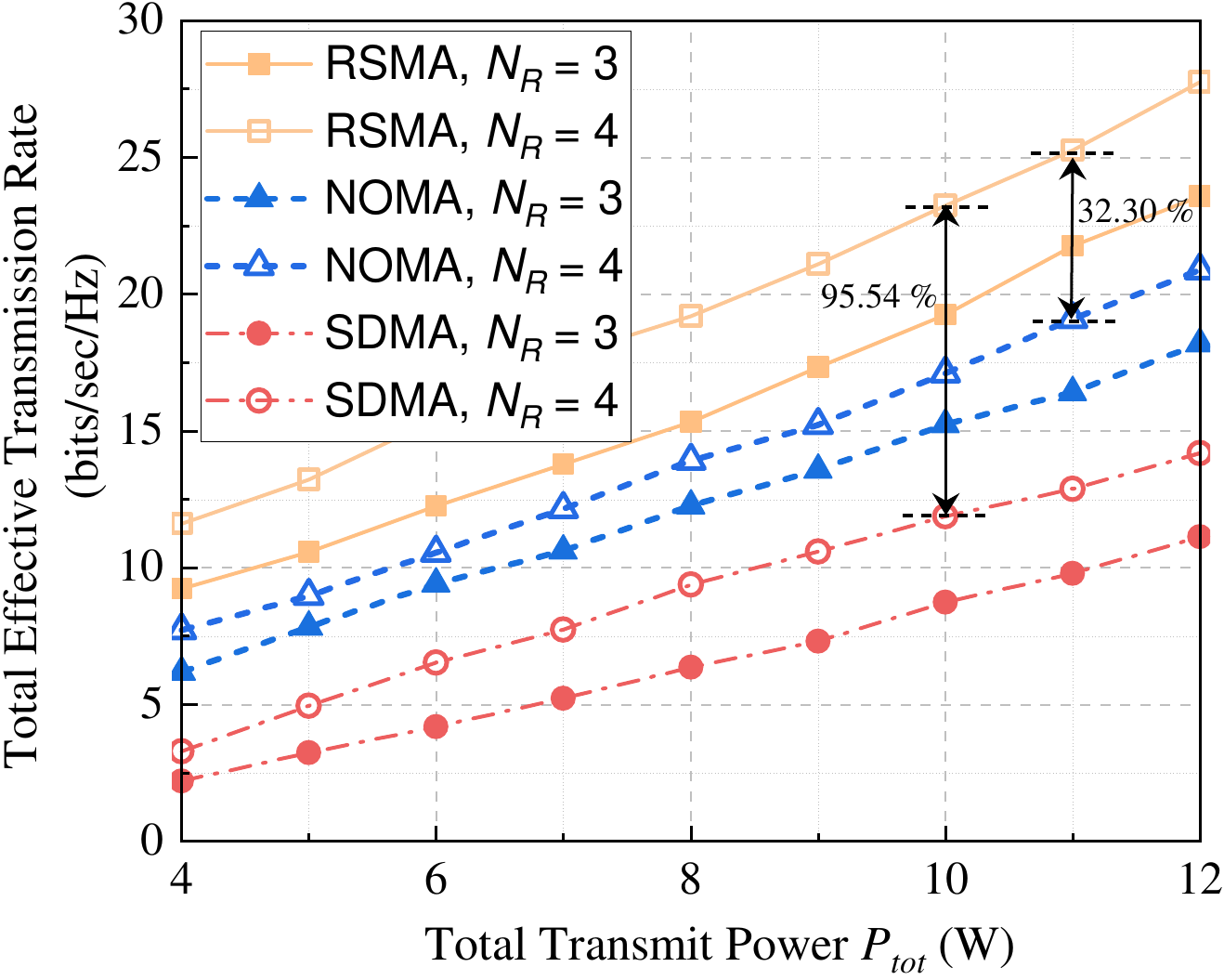}
    \caption{Relationship between the maximum total ETR and the total transmit power. $U = 5$ and $\mathcal{R}_{min} = 2.0$ bits/Sec/Hz.}
    \label{fig10}
  \end{minipage}
  \hfill
  \begin{minipage}[b]{0.3\textwidth}
    \includegraphics[scale=0.25]{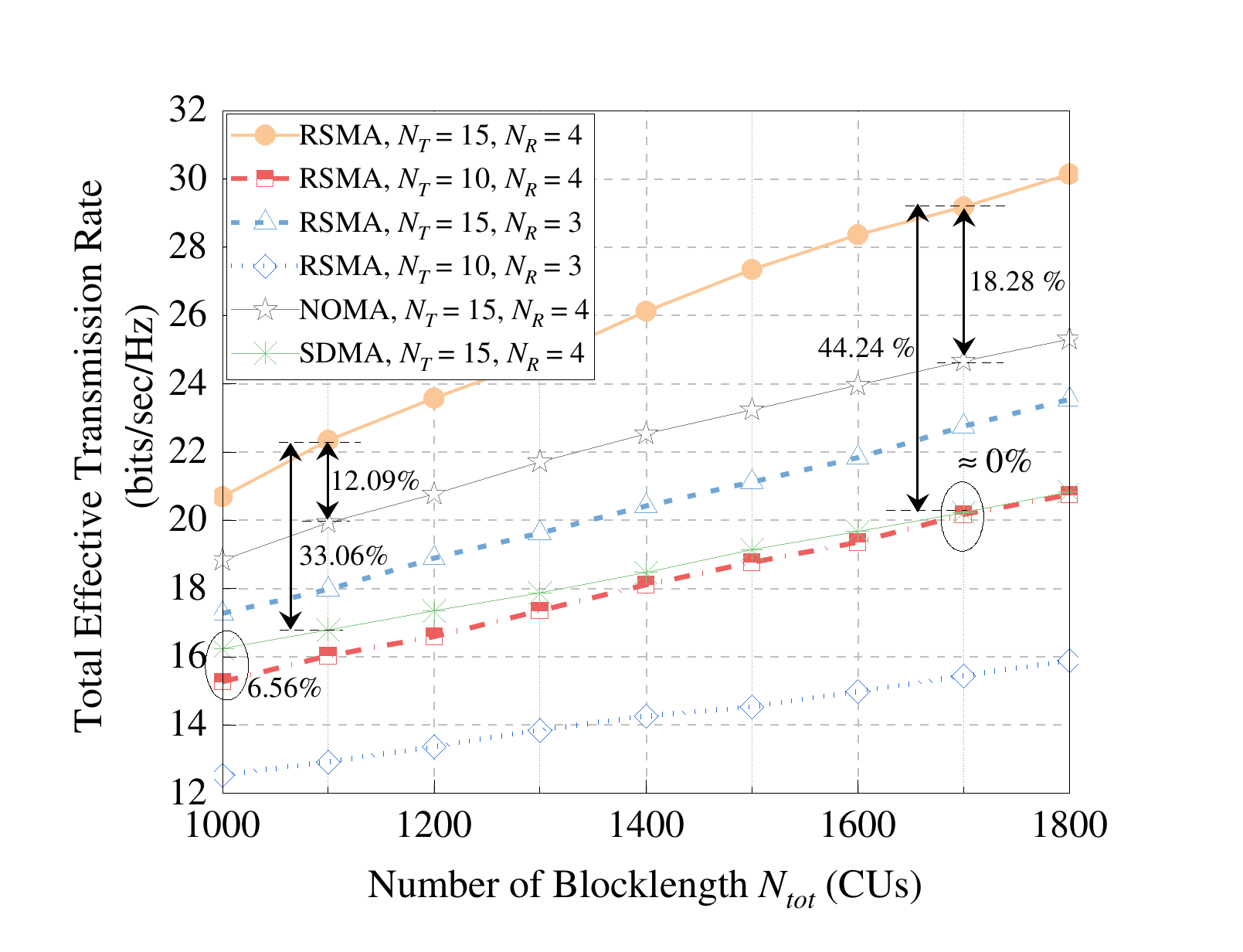}
   \caption{Relationship between the maximum total ETR and the total blocklength. $U = 5$ and $\mathcal{R}_{min} = 2.0$ bits/Sec/Hz.}
   \label{fig11}
  \end{minipage}
  \vspace{-1.5em}
\end{figure*}

\par In Fig. \ref{fig8}, we analyze the relationship between the maximum total ETR and the number of UEs $U$. Both our architecture and NOMA firstly showcase increased total ETR with $U$, followed by a decline. Particularly, at $\varepsilon_{th} \!=\! 10^{-5}$, our developed network architecture peaks at $U \!>\! 9$ before declining, while NOMA peaks at $U \!>\! 8$. With $\varepsilon_{th} \!=\! 10^{-6}$, our architecture's peak shifts at $U \!>\! 8$, whereas NOMA's peak shifts at $U \!>\! 7$. Compared to NOMA, our architecture achieves approximately $55.05\%$ and $44.74\%$ performance gains, respectively, primarily due to differences in SIC decoding mechanisms and interference management strategies. In NOMA, receivers prioritize decoding based on channel conditions. For the receiver with the $K$-th best channel condition, $U-K$ SIC operations need to be performed. Therefore, the DEPs of receivers with relatively better channel conditions will inevitably be affected by other receivers with relatively poorer channel conditions. Conversely, our architecture integrates RSMA, which only requires one SIC operation at each receiver, significantly reducing the adverse impacts of SIC operations on reliability. Compared with NOMA, our architecture better leverages diversity gains to accommodate more UEs, further demonstrating the RSMA's potential applications in the forthcoming xURLLC. Furthermore, as $U$ increases, the maximum total ETR achievable by SDMA decreases. This is mainly due to SDMA's reliance on spatial division to distinguish signals from different receivers, which can lead to interference between users as the density of UEs increases, thereby affecting transmission rates. Moreover, as wireless resources are inherently limited, an increase in $U$ leads to more frequent resource contention.

\vspace{-0.2em}

\par Fig. \ref{fig9} elucidates the relationship between the maximum total ETR and $\mathcal{R}_{min}$. It can be intuitively observed that RSMA outperforms NOMA and SDMA, exhibiting slower performance degradation rates. Specifically, when $\mathcal{R}_{min}$ increases from $2.0$ to $4.5$ bits/Sec/Hz, our architecture only experiences a $41.16\%$ loss in maximum total ETR, while NOMA and SDMA generate losses of approximately $62.99\%$ and $98.92\%$, respectively. These numerical results demonstrate the robustness of our architecture in maintaining the stability of transmission rates. This is mainly because RSMA can flexibly adjust the transmission rates of common and private messages through rate splitting to accommodate varied $\mathcal{R}_{min}$. In contrast, NOMA and SDMA are considered two different extreme interference management strategies, with the former completely decoding interference and the latter treating interference as noise. Therefore, NOMA and SDMA lack such adaptability and suffer more significant performance degradation with higher $\mathcal{R}_{min}$ due to their extreme interference management.

\par In Fig. \ref{fig10}, the relationship between the maximum total ETR and the total transmit power $P_{tot}$ is explored. As expected, the maximum total ETR increases with $P_{tot}$ for all schemes. Moreover, our developed network architecture outperforms NOMA and SDMA. For instance, at $N_{R} \!=\! 4$ and $P_{tot} \!=\! 10$ W, the performance gap between RSMA and SDMA is $95.54\%$. At $N_{R} \!=\! 4$ and $P_{tot} \!=\! 11$ W, the performance gains of RSMA exceed NOMA by $32.30\%$. This is primarily because increasing $P_{tot}$ can enhance the signal strength at receivers, thereby improving channel quality, which in turn reduces DEPs and improves the reliability of short-packet data transmission. Additionally, RSMA's strategy of partial interference decoding and treating interference as noise enables more effective interference management. In contrast, NOMA and SDMA employ simpler interference management strategies, limiting their performance gains.

\par Fig. \ref{fig11} illustrates the relationship between maximum total ETR and blocklength $N_{tot}$, with a target latency $D_{th} \!=\! 1$ \emph{ms} and available bandwidth $B_{tot}$ corresponding to $N_{tot} \!=\! 1000 \!\sim \!1800$ CUs, ranging from $B_{tot} \!=\! 1.0 \!\sim \! 1.8$ MHz. It can be observed that the maximum total ETRs of all three frameworks increase with the $N_{tot}$, which is consistent with \textbf{Corollary 1}. Numerical results reveal that our architecture outperforms NOMA and SDMA, with performance gains expanding as $N_{tot}$ increases. For instance, at $N_{tot} \!=\! 1100$ CUs, our architecture achieves approximately $12.09\%$ and $33.06\%$ performance gains compared to NOMA and SDMA, respectively. However, at $N_{tot} \!=\! 1700$ CUs, these improvements expand to $18.28\%$ and $44.24\%$, respectively. This fully demonstrates our architecture's ability to achieve higher total ETR performance with fewer resources under the identical target delay, which is beneficial for meeting stricter reliability constraints for xURLLC. Furthermore, the robustness of the developed network architecture can also be validated. Even with fewer transmit antennas, our scheme exhibits significant improvements in terms of maximum total ETR performance compared to SDMA.

\vspace{-0.3em}
\vspace{-0.2em}

\section{Conclusion And Future Outlook}

\vspace{-0.3em}

\par In this paper, we have embarked on a pioneering exploration for xURLLC's NGAT design and developed an innovative RSMA-mMIMO-xURLLC network architecture to accommodate the critical QoS constraints mandated by xURLLC under imperfect CSIT and FBL regimes. Subsequently, we have formulated a joint rate-splitting, beamforming, and transmit antenna selection optimization problem with the overarching objective of maximizing the total ETR. To effectively tackle this problem, we have reformulated and decomposed it into three subproblems by profoundly exploring the relationships between SINRs of common and private streams and antenna amounts, large-scale fading, transmit power, and blocklength. Moreover, we have proposed a low-complexity JPRT optimization algorithm to efficiently alternate the optimization of these three subproblems. Extensive simulations have substantiated the optimality and exceptional convergence performance of the proposed JPRT optimization algorithm. Compared with the state-of-the-art multi-access schemes, the extraordinary performance of the developed RSMA-mMIMO-xURLLC network architecture has been demonstrated.

\par In our forthcoming endeavors, we aim to incorporate statistical QoS provisioning mechanisms into the RSMA-mMIMO-xURLLC architecture developed in this paper. This entails a comprehensive analysis of the tail distribution for xURLLC's stringent QoS requirements, thereby endowing our network architecture with enhanced resilience under rare and extreme wireless scenarios.

\vspace{-0.3em}

\begin{appendices}
\section{Proof of Lemma 1}
\setcounter{equation}{0}
\renewcommand\theequation{A-\arabic{equation}}
\vspace{-0.3em}
  Given the random variable $R \in \left(R_{min},R_{max}\right)$, the p.d.f. of $R$ can be derived owing to the uniform distribution of UEs as follows:
  \vspace{-0.2em}
  \begin{equation}\label{A1}
     p_{R}\!\left(r\right) = \frac{2r}{R_{max}^{2} \! - \! R_{min}^{2}}.
  \end{equation}
  (i) Define $X_{1} \!=\! \frac{\kappa_{u}}{1 + N_{p}\rho_{p}\kappa_{u}}$, the c.d.f. of $X_{1}$ can be derived by
  \vspace{-0.4em}
  \begin{small}
   \begin{equation}\label{A2}
     \begin{aligned}
        P_{X_{1}}\!(x_{1}) & \! = \! \mathbb{P}\left\{X_{1} \! \leq \! x_{1}\right\} \! = \! \mathbb{P}\!\left\{\!\frac{\lambda^{2}}{(4\pi)^{2}(R^{2}\!+\!\tilde{h}^{2})\!+\!N_{p}\rho_{p}\lambda^{2}} \! \leq \! x_{1}\!\!\right\}\\
        & \overset{(a)}{=} 1 - \frac{1}{R_{max}^{2} \! - \! R_{min}^{2}}\left(\frac{\lambda^{2}\left(1 - N_{p}\rho_{p}x_{1}\right)}{(4\pi)^{2}x_{1}} - \tilde{h}^{2}\right),
     \end{aligned}
   \end{equation}
  \end{small}
  where $(a)$ can be derived from (\ref{A1}). Taking the derivative with respect to $P_{X_{1}}\!(x_{1})$, we have
  \vspace{-0.3em}
  \begin{small}
  \begin{equation}\label{A3}
     p_{X_{1}}\!(x_{1}) = \frac{\partial P_{X_{1}}\!(x_{1})}{\partial x_{1}} = \frac{\lambda^{2}}{(4\pi)^{2}\left(R_{max}^{2}-R_{min}^{2}\right)x_{1}^{2}}.
  \end{equation}
  \end{small}
  \vspace{-0.4em}
  In this case, we can obtain that
  \vspace{-0.3em}
  \begin{small}
   \begin{equation}\label{A4}
    \begin{aligned}
     & \widetilde{\Pi}_{1} = \mathbb{E}\left[\frac{\kappa_{u}^{2}}{\left(1 + N_{p}\rho_{p}\kappa_{u}\right)^{2}}\right] = \mathbb{E}\left[X_{1}^{2}\right] = \int_{X_{min}^{(1)}}^{X_{max}^{(1)}} x_{1}^{2}p_{X_{1}}\!(x_{1})\d x_{1}\\
     & = \frac{\lambda^{2}}{(4\pi)^{2}\!\left(R_{max}^{2}\!-\!R_{min}^{2}\right)}\!\int_{X_{min}^{(1)}}^{X_{max}^{(1)}} \!\! \mathrm{\mathbf{1}} \d x_{1} \!=\! \frac{\lambda^{2}\!\left(\!X_{max}^{(1)} \!-\! X_{min}^{(1)}\!\right)}{(4\pi)^{2}\!\left(R_{max}^{2}\!-\!R_{min}^{2}\right)}.
    \end{aligned}
   \end{equation}
  \end{small}
  \vspace{-0.4em}
  Similarly, we can also obtain that
  \vspace{-0.3em}
  \begin{small}
  \begin{equation}\label{A5}
     \!\!\!\!\!\! \widetilde{\Pi}_{2} = \mathbb{E}\!\!\left[\!\frac{\kappa_{u}^{3/2}}{1 + N_{p}\rho_{p}\kappa_{u}}\right] \!\approx\! \mathbb{E}\!\left[X_{1}\right] \!=\! \frac{\lambda^{2}\!\log\!\!\left(\frac{X_{max}^{(1)}}{X_{min}^{(1)}}\!\right)}{(4\pi)^{2}\!\left(R_{max}^{2}\!-\!R_{min}^{2}\right)},
  \end{equation}
  \end{small}
  \vspace{-0.2em}
  where $X_{min}^{(1)}$ and $X_{max}^{(1)}$ are given by \textbf{Lemma 1}.

  (ii) We define $X_{2}$ as follows:

   \vspace{-1em}

  \begin{equation}\label{A6}
      X_{2} \!=\! \frac{\kappa_{u}^{3}}{\left(1\!+\!N_{p}\rho_{p}\kappa_{u}\right)^{2}} \!=\! \left\{\!\!\!
                                                                                    \begin{array}{ll}
                                                                                      \frac{\kappa_{u}}{\left(N_{p}\rho_{p}\right)^{2}}, \!\!\!\!\! & \hbox{$N_{p}\rho_{p}\kappa_{u} \gg 1$;} \\
                                                                                      \kappa_{u}^{3}, \!\!\!\!\! & \hbox{$0<N_{p}\rho_{p}\kappa_{u}\ll 1$;} \\
                                                                                      \frac{\kappa_{u}^{3}}{4}, \!\!\!\!\! & \hbox{$N_{p}\rho_{p}\kappa_{u} \approx 1$.}
                                                                                    \end{array}
                                                                                  \right.
  \end{equation}
   When $N_{p}\rho_{p}\kappa_{u} \gg 1$, we can obtain that
   \vspace{-0.2em}
  \begin{equation}\label{A7}
     P_{X_{2}}\!\!\left(x_{2}\right) \overset{(a)}{=} 1 \!-\! \frac{1}{R_{max}^{2} \!-\! R_{min}^{2}}\!\left(\!\frac{\lambda^{2}}{(4\pi N_{p}\rho_{p})^{2}x_{2}} \!-\! \tilde{h}^{2}\!\right).
  \end{equation}
  \vspace{-0.2em}
  Taking the derivative with respect to $P_{X_{2}}\!\left(x_{2}\right)$, the p.d.f. of $X_{2}$ is given by
  \vspace{-0.2em}
  \begin{equation}\label{A8}
     p_{X_{2}}\!(x_{2}) = \frac{\lambda^{2}}{\left(R_{max}^{2}-R_{min}^{2}\right)\left(4\pi N_{p}\rho_{p}\right)^{2}x_{2}^{2}}.
  \end{equation}
  \vspace{-0.2em}
  As a result, we can obtain that
  \vspace{-0.4em}
  \begin{small}
  \begin{equation}\label{A9}
     \begin{aligned}
     \widetilde{\Pi}_{3} & = \frac{\lambda^{2}}{\left(4\pi N_{p}\rho_{p}\right)^{2}\!\left(R_{max}^{2} \!-\! R_{min}^{2}\right)}\log\!\!\left(\!\frac{R_{max}^{2} + \tilde{h}^{2}}{R_{min}^{2} + \tilde{h}^{2}}\!\right)\\
      & \overset{(b)}{\approx} \frac{\lambda^{2}}{8\left(\pi N_{p}\rho_{p}\right)^{2}\!\left(R_{max}^{2} \!-\! R_{min}^{2}\right)}\log\!\!\left(\!\frac{R_{max}}{R_{min}}\!\right),
     \end{aligned}
  \end{equation}
  \end{small}
  \vspace{-0.2em}
  where $(b)$ indicates that the height of the BS is much smaller than the radius of the cell.
  \vspace{-0.2em}

  When $0 < N_{p}\rho_{p}\kappa_{u} \ll 1$, the c.d.f. of $X_{2}$ can be similarly derived as follows:
  \vspace{-0.3em}
  \begin{small}
  \begin{equation}\label{A10}
     P_{X_{2}}\left(x_{2}\right) = 1 - \frac{1}{R_{max}^{2} - R_{min}^{2}}\!\left(\!\frac{\lambda^{2}}{(4\pi)^{2}x_{2}^{1/3}}+\tilde{h}^{2}\!\right).
  \end{equation}
  \end{small}

  By taking the derivative with respect to $P_{X_{2}}\left(x_{2}\right)$, we have

  \vspace{-0.8em}

  \begin{equation}\label{A11}
     p_{X_{2}}\!(x_{2}) = \frac{\lambda^{2}}{3(4\pi)^{2}\left(R_{max}^{2}-R_{min}^{2}\right)}x_{2}^{-4/3}.
  \end{equation}

  \vspace{-0.3em}

  In this case, we can obtain that
  \begin{small}
  \begin{equation}\label{A12}
     \!\!\!\!\!\!\!\!\!\!\!\! \widetilde{\Pi}_{3} \!=\! \frac{\lambda^{6}\!\!\left[\!\frac{1}{\left(R_{min}^{2}+\tilde{h}^{2}\right)^{2}}\!-\!\frac{1}{\left(R_{max}^{2}+\tilde{h}^{2}\right)^{2}}\!\right]}{2(4\pi)^{6}\!\left(R_{max}^{2} \!-\! R_{min}^{2}\right)} \!\overset{(b)}{\approx} \! \frac{\lambda^{6}\!\!\left(R_{max}^{2} \!+\! R_{min}^{2}\right)}{2(4\pi)^{6}R_{max}^{4}R_{min}^{4}}.\!\!\!\!
  \end{equation}
  \end{small}

  Finally, when $N_{p}\rho_{p}\kappa_{u} \approx 1$, according to (\ref{A6}) and (\ref{A12}), we can directly derive that
  \begin{equation}\label{A13}
    \widetilde{\Pi}_{3} \approx \frac{\lambda^{6}\!\!\left(R_{max}^{2} \!+\! R_{min}^{2}\right)}{8(4\pi)^{6}R_{max}^{4}R_{min}^{4}}.
  \end{equation}
  Combining (\ref{A9}), (\ref{A12}), and (\ref{A13}), Eq. (\ref{e25}) can be ultimately derived.
%  \begin{equation}\label{A14}
%     \widetilde{\Pi}_{3} \!=\! \mathbb{E}\!\!\left[\!\frac{\kappa_{u}^{3}}{\left(1\!+\!N_{p}\rho_{p}\kappa_{u}\right)^{2}}\!\right] \!=\! \left\{\!\!\!\!
%                                                                                                                        \begin{array}{ll}
%                                                                                                                          \frac{\lambda^{2}\log\left(\frac{R_{max}}{R_{min}}\right)}{8\left(\! \pi N_{p}\rho_{p}\!\right)^{2}\left(R_{max}^{2} - R_{min}^{2}\right)}, \hbox{$N_{p}\rho_{p}\kappa_{u} \!\gg\! 1$;} \\
%                                                                                                                          \frac{\lambda^{6}\!\left(R_{max}^{2} \!+\! R_{min}^{2}\right)}{2(4\pi)^{6}R_{max}^{4}R_{min}^{4}}, \hbox{$0 < N_{p}\rho_{p}\kappa_{u} \!\ll\! 1$;} \\
%                                                                                                                          \frac{\lambda^{6}\!\left(R_{max}^{2} \!+\! R_{min}^{2}\right)}{8(4\pi)^{6}R_{max}^{4}R_{min}^{4}}, \hbox{$N_{p}\rho_{p}\kappa_{u} \!\approx\! 1$.}
%                                                                                                                        \end{array}
%                                                                                                                      \right.
%  \end{equation}

  (iii) Following a similar derivation process as in (\ref{A6})-(\ref{A13}), $\widetilde{\Pi}_{4}$ and $\widetilde{\Pi}_{5}$ can be easily derived, and their closed-form expressions have been presented in (\ref{e26}) and (\ref{e27}), respectively.

\vspace{-1em}

\section{Proof of Lemma 2}
\setcounter{equation}{0}
\renewcommand\theequation{B-\arabic{equation}}

\vspace{-0.5em}

According to \cite{sun2018short, 10382447}, we can directly prove that $\varepsilon_{c,u}\big(N_{d},\boldsymbol{\rho},\mathcal{R}_{c}\big)$ and $\varepsilon_{p,u}\big(N_{d},\boldsymbol{\rho},\mathcal{R}_{p,u}\big)$ are strictly monotone decreasing functions with respect to $\rho_{c}$ and $\rho_{p,u}$, respectively. Next, we prove that $\varepsilon_{c,u}\big(N_{d},\boldsymbol{\rho},\mathcal{R}_{c}\big)$ is a strictly monotone decreasing function with respect to $N_{d}$. Taking partial derivatives w.r.t. $N_{d}$ for $\varepsilon_{c,u}\big(N_{d},\boldsymbol{\rho},\mathcal{R}_{c}\big)$, we can derive that

\vspace{-1em}

\begin{equation}\label{B1}
   \begin{aligned}
      & \frac{\partial}{\partial N_{d}} \varepsilon_{c,u}\big(N_{d},\boldsymbol{\rho},\mathcal{R}_{c}\big) = \frac{\partial}{\partial N_{d}} Q\big(g(N_{d},\boldsymbol{\rho},\mathcal{R}_{c})\big)\\
      & = -\frac{N_{d}^{-\frac{1}{2}}}{2\sqrt{2\pi}}e^{-\frac{g^{2}(N_{d},\boldsymbol{\rho},\mathcal{R}_{c})}{2}} \! \cdot \!\! \left(\!\frac{\sum\limits_{j=1}^{N_{R}} \! \ln\!\left(\!1\!+\!\Gamma_{c,u}^{(j)}\!\right)\!-\!\mathcal{R}_{c}\ln 2}{\sum\limits_{j=1}^{N_{R}}\!\!\sqrt{\mathcal{V}(\Gamma_{c,u}^{(j)})}}\!\!\right) < 0.
   \end{aligned}
\end{equation}

\par From (\ref{B1}), we can derive that $\varepsilon_{c,u}\big(N_{d},\boldsymbol{\rho},\mathcal{R}_{c}\big)$ strictly decreases with $N_{d}$. Similarly, it can be obtained that $\varepsilon_{p,u}\big(N_{d},\boldsymbol{\rho},\mathcal{R}_{p,u}\big)$ also monotonically decreases with $N_{d}$.

\par Finally, we show that $\varepsilon_{c,u}\big(N_{d},\boldsymbol{\rho},\mathcal{R}_{c}\big)$ is monotonically increasing with $\mathcal{R}_{c}$. Taking partial derivatives w.r.t. $\mathcal{R}_{c}$ for $\varepsilon_{c,u}\big(N_{d},\boldsymbol{\rho},\mathcal{R}_{c}\big)$, we can derive that

\vspace{-1.2em}

\begin{equation}\label{B2}
   \frac{\partial}{\partial \mathcal{R}_{c}} \varepsilon_{c,u}\big(N_{d},\boldsymbol{\rho},\mathcal{R}_{c}\big) = \frac{\ln 2 \cdot e^{-\frac{g^{2}(N_{d},\boldsymbol{\rho},\mathcal{R}_{c})}{2}} }{\sqrt{2\pi\sum\limits_{j=1}^{N_{R}}\mathcal{V}(\Gamma_{c,u}^{(j)})}} > 0.
\end{equation}

\vspace{-0.5em}

\par From (\ref{B2}), we can derive that $\varepsilon_{c,u}\big(N_{d},\boldsymbol{\rho},\mathcal{R}_{c}\big)$ strictly increases with $\mathcal{R}_{c}$. Similarly, it can be obtained that $\varepsilon_{p,u}\big(N_{d},\boldsymbol{\rho},\mathcal{R}_{p,u}\big)$ also monotonically decreases with $\mathcal{R}_{p,u}$.

\section{Proof of Corollary 1}
\setcounter{equation}{0}
\renewcommand\theequation{C-\arabic{equation}}

\par We denote the optimal solution of $\mathcal{P}2$ by $\left\{N_{p}^{\dag},N_{d}^{\dag},\boldsymbol{\mathcal{R}}^{\dag},\boldsymbol{\rho}^{\dag}\right\}$. Then, we employ proof by contradiction to show that (\ref{e38b}) always holds. If $\varepsilon_{c,u}\big(N_{d}^{\dag},\boldsymbol{\rho}^{\dag},\mathcal{R}_{c}^{\dag}\big) < \varepsilon_{th}$, based on \textbf{Lemma 2}, we can increase $\mathcal{R}_{c}^{\dag}$ to $\mathcal{R}_{c}^{\ast}$ until $\varepsilon_{c,u}\big(N_{d}^{\dag},\boldsymbol{\rho}^{\dag},\mathcal{R}_{c}^{\ast}\big) \!=\! \varepsilon_{th}$ without violating any other constraints, which contradicts the optimality assumption of $\mathcal{R}_{c}^{\dag}$. As a result, we can conclude that $\varepsilon_{c,u}\big(N_{d},\boldsymbol{\rho},\mathcal{R}_{c}\big) = \varepsilon_{th}$ can be always guaranteed at the optimal solutions. Similarly, we can prove that constraint (\ref{e38c}) is active at the optimal solutions. For constraint (\ref{e23b}), using proof by contradiction, let's assume $N_{p}^{\dag} > N_{T}$ holds at the optimal solutions. Then, we can decrease $N_{p}^{\dag}$ to $N_{p}^{\ast} \!=\! N_{T}$, and increase $\rho_{p}^{\dag}$ to $\rho_{p}^{\ast} \!=\! q \rho_{p}^{\dag}$ with $q \!=\! \frac{N_{p}^{\dag}}{N_{T}} \!>\! 1$ such that $N_{p}^{\ast} \rho_{p}^{\ast} \!=\! N_{p}^{\dag} \rho_{p}^{\dag}$. It is worth noting that $N_{d}$ and $\rho_{p}$ always appear as a product $N_{p} \rho_{p}$ in the power constraint (\ref{e23h}). In this case, we have $\mathcal{R}_{c}^{\ast} \!=\! \mathcal{R}_{c}^{\dag}$ without violating any other constraints, where $\mathcal{R}_{c}^{\ast}$ is a new transmission rate of common stream under $N_{p}^{\ast}$ and $\rho_{p}^{\ast}$. This contradicts the optimality of $\mathcal{R}_{c}^{\dag}$ and $N_{p}^{\dag} \!>\! N_{T}$. Therefore, it can be concluded that there exists at least one optimal solution with $N_{p} \!=\! N_{T}$.

\vspace{-1.5em}

\section{Proof of Lemma 3}
\setcounter{equation}{0}
\renewcommand\theequation{D-\arabic{equation}}

\vspace{-0.5em}

\par To prove the monotonic increasing property of $g(N_{d},\boldsymbol{\Gamma},R)$ w.r.t. $\boldsymbol{\Gamma}$, as per (\ref{e2}), it suffices to find the first-order partial derivative of $g(N_{d},\boldsymbol{\Gamma},R)$ w.r.t. $\boldsymbol{\Gamma}$, denoted as $\frac{\partial g}{\partial \boldsymbol{\Gamma}} > 0$. For the concave property of $g(N_{d},\boldsymbol{\Gamma},R)$, we seek the second-order partial derivative of $g(N_{d},\boldsymbol{\Gamma},R)$ w.r.t. $\boldsymbol{\Gamma}$, denoted as $\frac{\partial^{2} g}{\partial \boldsymbol{\Gamma}^{2}} < 0$. The detailed proof process involves only basic rules of differentiation, which readers can validate using tools like \texttt{Wolfram Mathematica}. We omit these steps here. % We have plotted the graph of $g(N_{d},\boldsymbol{\Gamma},R)$ to intuitively illustrate Lemma 3, as depicted in Fig. 3.

\section{Proof of Lemma 4}
\setcounter{equation}{0}
\renewcommand\theequation{E-\arabic{equation}}
\par According to (\ref{e48}) and (\ref{e49}), for any $u \in \mathcal{U}$, the constraint $\varepsilon_{p,u}(\boldsymbol{\rho},\mathcal{R}_{p,u}) \leq \varepsilon_{th}$ can be satisfied in $\left(0,\mathcal{R}_{p,u}^{\uparrow}\right)$, which is equivalent to $0 < Q^{-1}(\varepsilon_{th}) < g\left(N_{d},\boldsymbol{\Gamma},\mathcal{R}_{u,p}\right)$. In this case, by taking the second-order partial derivative of $\widetilde{\mathcal{R}}_{p,u}$ with respect to $\mathcal{R}_{p,u}$, we can derive that

\vspace{-1.0em}

\begin{equation}\label{E1}
   \frac{\partial^{2} \widetilde{\mathcal{R}}_{p,u}}{\partial \mathcal{R}_{p,u}^{2}} \!=\! -\frac{F_{1}\!\left(\mathcal{R}_{p,u}\right)}{\sqrt{2\pi} \mathcal{V}\!\left(\boldsymbol{\Gamma}_{p,u}\right)}\cdot \exp\!\!\left\{\!-\frac{g^{2}\!\left(N_{d},\boldsymbol{\Gamma}_{p,u},\mathcal{R}_{p,u}\right)}{2}\!\right\} \!<\! 0,
\end{equation}

\vspace{-1.0em}

where
\begin{equation}\label{E2}
   F_{1}\!\left(\mathcal{R}_{p,u}\right) \!=\! (2\ln2)\widetilde{\mathcal{V}}\!\left(\Gamma_{p,u}\right) + (\ln 2)^{2}\mathcal{R}_{p,u}N_{d}g\!\left(N_{d},\boldsymbol{\rho},\mathcal{R}_{p,u}\right) \!>\! 0,
\end{equation}
in which $\widetilde{\mathcal{V}}\left(\Gamma_{p,u}\right) = \sqrt{N_{d}\sum\limits_{j=1}^{N_{R}}\mathcal{V}\left(\Gamma_{p,u}^{(j)}\right)}$. Therefore, $\widetilde{\mathcal{R}}_{p,u}$ is a strictly convex function w.r.t. $\mathcal{R}_{p,u}$. So, \textbf{Lemma 4} can be concluded.

\vspace{-0.5em}

\section{Proof of Lemma 5}
\setcounter{equation}{0}
\renewcommand\theequation{F-\arabic{equation}}
\par Similarly, for any $u \in \mathcal{U}$, the constraint $\varepsilon_{c,u}(\boldsymbol{\rho},\mathcal{R}_{c,u}) \leq \varepsilon_{th}$ can be satisfied in $\left(0,\mathcal{R}_{c}^{\uparrow}\right)$, which is equivalent to $0 < Q^{-1}(\varepsilon_{th}) < g\left(N_{d},\boldsymbol{\Gamma},\mathcal{R}_{c}\right)$. As a result, the second-order partial derivative of $\widetilde{\mathcal{R}}_{c,u}$ with respect to $\mathcal{R}_{c,u}$ can be derived as follows:

\vspace{-0.8em}

\begin{small}
\begin{equation}\label{F1}
   \begin{aligned}
      & \frac{\partial^{2} \widetilde{\mathcal{R}}_{c}}{\partial \mathcal{R}_{c}} = -\sum\limits_{k = 1, k \neq u} \!\!\frac{F_{2}\left(\mathcal{R}_{c}\right)}{\sqrt{2\pi} \mathcal{V}\left(\boldsymbol{\Gamma}_{c,k}\right)} \cdot \exp\left\{-\frac{g^{2}\left(\Gamma_{c,k},N_{d},\mathcal{R}_{c}\right)}{2}\right\}\\
      & - \frac{\ln2 \sqrt{N_{d}}F_{3}\left(\mathcal{R}_{c}\right)}{\sqrt{2\pi} \mathcal{V}\left(\Gamma_{c,u}\right)}\cdot \exp\left\{-\frac{g^{2}\left(\Gamma_{c,u},N_{d},\mathcal{R}_{c}\right)}{2}\right\},
   \end{aligned}
\end{equation}
\end{small}

\vspace{-1em}

where

\vspace{-1.5em}

\begin{equation*}\label{F2}
    \left\{
      \begin{array}{ll}
       \!\!\! F_{2}\!\left(\mathcal{R}_{c}\right) \!=\! \left(\ln 2\right)^{2}\!N_{d}\mathcal{R}_{c,k}^{min}g\!\left(\Gamma_{c,k},N_{d},\mathcal{R}_{c}\right) \!>\! 0, \\
       \!\!\! F_{3}\!\left(\mathcal{R}_{c}\right) \!=\! 2\sqrt{\widetilde{\mathcal{V}}\!\left(\Gamma_{c,k}\right)} \!+\! \ln2 \sqrt{N_{d}} \mathcal{R}_{c,k}^{max}g\!\left(\Gamma_{c,k},N_{d},\mathcal{R}_{c}\right) \!>\! 0,
      \end{array}
    \right.
\end{equation*}

\vspace{-0.8em}

\par Thus, we can derive that $\frac{\partial^{2} \widetilde{\mathcal{R}}_{c}}{\partial \mathcal{R}_{c}}<0$, which implies that $\widetilde{\mathcal{R}}_{c}$ is a strictly concave function w.r.t. $\mathcal{R}_{c}$ in $\left(0,\mathcal{R}_{c}^{\uparrow}\right)$.
\end{appendices}
\vspace{-0.5em}
\footnotesize
\bibliographystyle{IEEEtran}
\bibliography{IEEEabrv,reference_globecom}
\end{document}